\documentstyle[12pt]{article}
\def\beq{\begin{equation}}
\def\eeq{\end{equation}}
\newcommand{\bea}{\begin{eqnarray}}
\newcommand{\eea}{\end{eqnarray}}
\newcommand{\nn}{\nonumber}

\def\R{ {\rm R \kern -.31cm I \kern .15cm}}
\def\C{ {\rm C \kern -.15cm \vrule width.5pt \kern .12cm}}
\def\Z{ {\rm Z \kern -.27cm \angle \kern .02cm}}
\def\N{ {\rm N \kern -.26cm \vrule width.4pt \kern .10cm}}
\def\1{{\rm 1\mskip-4.5mu l} }
\def\lsim{\raise0.3ex\hbox{$<$\kern-0.75em\raise-1.1ex\hbox{$\sim$}}}
\begin{document}

\begin{titlepage}
\begin{flushright} FANSE-96/19\\LPTHE Orsay-96/108\\ hep-ph/9702254
\end{flushright}

\begin{centering}

{\large \bf CP-violating Reflection of High Energy Fermions during a
 first order Phase Transition}
\bigskip
\bigskip

{\large J. Rodr\'{\i}guez-Quintero$^{a,}$\footnote{e-mail: Jquinter@cica.es}, 
O. P\`{e}ne$^{b,}$\footnote{e-mail: Pene@qdc.th.u-psud.fr}, 
M. Lozano$^{a,}$\footnote{e-mail: Lozano@cica.es} }

\bigskip

\noindent
$^a$ \ Dpto. de F\'{\i}sica At\'omica, Molecular y Nuclear, Universidad de 
Sevilla, Spain\footnote{Work partially supported by Spanish CICYT, project 
PB 95-0533-A}\\
\noindent
$^b$ \ LPTHE, F 91405 Orsay, France\footnote{Laboratoire associ\'e au
Centre National 
de la Recherche Scientifique, URA  D0063.}\\

\end{centering}

\bigskip

\begin{abstract}

We study the high energy behaviour of fermions hitting a general wall 
caused by a first-order phase transition. The wall profile is 
introduced through general analytic and non-analytic functions. The 
reflection coefficient is computed in the high energy limit and its 
connection with the analytic properties of the wall profile function 
is shown. The high energy behaviour of the fermions hitting the wall 
is determined either by the leading singularity, i.e. the closest pole 
to the real axis, when the profile function is analytic, or by the first 
non-continuous derivative on the real axis, in the non-analytic case. 
CP-violating wall profiles are studied and it is shown that the respective  
symmetry properties of the CP-conserving and CP-violating profile functions
plays an important role on the size of the CP asymmetry.

\bigskip
\bigskip

\noindent PACS. \ :

\begin{itemize}

\item{11.10Q-}Field theory, relativistic wave equations.
\item{11.80F-}Relativistic scattering theory, approximations.
\item{11.30Q-}Symmetry and conservation laws, spontaneous symmetry breaking.

\end{itemize}
\end{abstract}
\medskip

\end{titlepage}

\section{Introduction}

\bigskip

In ref. \cite{Rod96} we studied the problem of transmission and reflection of 
fermions through a wall and established the relationship which connects the 
complex plane poles of the wall profile function with the high energy 
reflection coefficient. The aim of the present paper is to study in more details
this previous result and to go further into its consequences, mainly CP violation.
The interest in the above-mentioned old problem has been renewed in 
the last years by the proposal that the baryon asymmetry of the universe 
might have been produced if a first order $SU(2) \times U(1)$ phase transition 
took place during the cosmological evolution. This electroweak phase transition 
can be described in terms of bubbles of ``true'' vacuum with an inner 
expectation value of the Higgs field $v \not= 0$, i.e. a spontaneously 
non-symmetric phase, appearing and expanding in the preexisting ``false'' 
vacuum with $v = 0$. If the weak phase transition is
first order the standard model of weak interactions has qualitatively
all the ingredients to 
produce the cosmological baryon asymmetry \cite{[1]}, i.e. baryon number 
violation, C and CP-violation and out-of-equilibrium processes
\cite{[2]}. The question is whether it can produce a large enough
asymmetry to account for the observed baryon number of the universe:
$n_B/s \sim (4-6) 10^{-11}$.
CP-violation in the Kobayashi-Maskawa scheme \cite{[3]} has been claimed to be
large enough to generate alone the observed baryon asymmetry \cite{[4]}. 
However a zero temperature estimate \cite{[5]} of
 the GIM suppression of electroweak C and CP effects, as well as a
finite temperature 
study stressing the effect of gluonic thermal fluctuation \cite{Quimb}, allows
to argue that the SM scenario \cite{[4]} produces a CP asymmetry more than
 ten order of magnitudes too small, $n_B/s \lsim  10^{-20}$.

 Simple extensions of the SM may produce much larger CP asymmetries, such  
as the two-Higgs-doublet \cite{[6]}, \cite{joyce}, \cite{comelli} and
\cite{cline}, left-right symmetric models \cite{frere},  SUSY
\cite{susy} or models with heavy leptons \cite{[10]}, \cite{pilar}. These 
models  contain several Higgs fields and a relative phase between the
Higgs vacuum expectation values may generate a new  source of CP
violation. The latter CP violation may be much larger than in the
standard model for two reasons: i) CP violation often appears at the {\it tree
level} while in SM it is a loop effect; ii) there is more freedom in
the parameters than in the SM. Even so, the parameters in the model of
ref. \cite{pilar}  are too constrained to allow for enough CP violation,
but the discrepancy is only two orders of magnitudes as compared to the 
ten orders of magnitude of the SM.

We will therefore concentrate on tree level CP violation due
to vacuum expectation values of Higgs fields beyond the standard model.
These CP-violating processes do not directly violate baryon number nor 
fermion number. But they can create local inhomogeneities in the 
baryon (lepton) number, for example when the
fermions in the plasma hit a bubble wall: an excess of antibaryons (baryons)
 may be created outside (inside) the bubble, which may subsequently be converted 
\cite{[14]}, via anomalous weak interaction\cite{Hoo76}, into a baryon asymmetry
in the universe. 

We thus need to study the
problem of particles, propagating in the plasma, say from outside the
bubble, penetrating the wall and being eventually reflected outside or
transmitted inside the bubble.
This is a difficult problem since the fermions  interact not only with
the wall (i.e. the Higgs vacuum expectation value) but also with the 
particles in the surrounding plasma. It involves solving a quantum Fokker-Planck 
equation taking into account CP-violation and baryon anomaly. A useful 
simplifying assumption has been proposed, the so-called 
``charge transport'' mechanism \cite{[6]}, in which the anomalous weak 
interaction happens at a distance from the bubble wall. This assumption consists
in decomposing the process into two steps, one describing the production of the 
CP asymmetry when the quarks/antiquarks are reflected/transmitted on
the wall, the second
describing the transport and eventual transformation of this CP asymmetry via 
the baryon number anomaly. We will concentrate on the first step, the
interaction of the incoming fermion with the wall. 
It may be assumed that during this first step the diffusion 
corrections are relatively minor ones to the scattering from the bubble
wall, which is driven by the Higgs expectation value (during the second
step they are, of course, crucial). The
thermal effects are {\it only} considered via the introduction of the
Higgs field effective potential, which
takes into account the thermal bath temperature. For simplicity we will
thus describe the reflection/transmission mechanism  of fermions at
zero temperature on a wall generated by a Higgs expectation value which
depends on $\vec x$. The thermal effects could also be present via  the
thermal modification of the fermion spectrum (thermal masses). We will
not consider them, although we believe that our conclusions in the
present paper can easily be extended to incorporate them. 
Furthermore, the information given for the solution of the academic
$T=0$ case can also be of use in other problems of physics, in solid
state physics, in astrophysics, etc.

The study of the reflection and transmission of fermions on a bubble
surface is thus 
crucial to elucidate whether it may produce a large enough CP-asymmetry 
to generate
the correct baryon asymmetry of the universe. The size of the bubble is
in general very large compared to the typical size of the CP violating
process under study. Therefore we will approximate the bubble surface
by a flat wall perpendicular to the $z$ direction. The Higgs
expectation values depend on  $z$, and the resulting effective mass of
the fermion is a function of $z$. This function depends on the  
above mentioned Higgs field effective potential. The 
profile obtained by solving the equation of motion is therefore rather complex 
and depends on many coupling constants \cite{[6]}.

A frequently used simplifying assumption
is to represent the wall profile by a step function  considering an 
extremely sharp phase transition which allows to compute, in an exact way, the 
two-point Green function
for a free fermion in the presence of this thin wall \cite{[5]}. 
However, it is known and will clearly appear later in this paper that
{\it in such a 
 thin wall approximation all CP asymmetries due to Higgs
expectation value at the tree level do vanish}. The CP violating phase
can just be rotated away by a chiral rotation. Therefore {\it we will
now concentrate on thicker walls},
although the wall thickness must be
smaller enough than the mean free path of the fermions in the plasma to neglect 
the diffusion corrections.

In general, 
for more complex wall profiles the task of obtaining the two-point Green function 
is very difficult and analytic results for this Green function may not exist 
in most cases or be in practice unreachable. One may try numerical 
calculations, but, since Green functions are not really necessary at tree level
we will not insist in this direction. Still, the first step is to find
an orthonormal and complete set of eigenfunctions. The orthogonality of these 
eigenfunctions is a crucial question which can be considered in a general way. 
In appendix B, we show the orthogonality of the two different types of eigenstates 
built by requiring appropriate asymptotic behaviours far from the wall as in 
refs. \cite{[7]}, \cite{[8]}.
One describes the wall  by an ansatz that simulates the dynamics of the 
phase transition \cite{[6]} \cite{[7]} \cite{[8]}. There exists one such ansatz for which, in 
refs. \cite{[7]} and \cite{[8]}, an analytical solution is given and a numerical 
calculation has been performed in \cite{[6]}. In both cases a new net effect of
CP-violation is obtained at tree level, that cannot 
be obtained for the thin wall.

It is known that observable CP asymmetries can only be obtained if,
besides a CP violating (CP-odd) phase, a CP-even phase (a phase which
is equal for particles and anti-particles) exists and interferes with
the CP-odd one. 
In the 
thin wall approximation a CP-even phase only appears in the total reflection
domain.
This is no longer true in the thick wall case, and CP-violating effects 
may exist outside the
total reflection region, up to infinite energy of the incoming fermion.

 A completely general solution of the thick wall problem is difficult.
However, there is a domain in which a few
{\it exact results can be derived: when the energy of the incoming
fermion is large.}
In the present paper we develop a general
method to calculate
in the {\it high energy limit} the reflection coefficient of fermions
hitting a wall,
establishing a simple relationship between the analytic properties of the profile
function and the high energy behaviour. The quantum correction to the
expected classical
behaviour is obtained and its importance depending on the above mentioned 
analytic properties is
showed. As we will point out, the range of energy for the quarks in the
electroweak phase
transition is appropriate to consider this limit, except for the top. 
The CP-violating
effects are incorporated by means of an appropriate imaginary mass
term and the basic quantity  $|R(E)|^2 - |\overline{R}(E)|^2$  is
obtained. Interesting general theoretical consequences for the fermions
quantum scattering and for the CP-violating effects will be derived from
this connection
of the analytic properties of the profile function and the high energy
behaviour. Apart of the clear interest of this problem in cosmology, it is
possible to apply
the formalism to systems of relativistic fermions that can suffer a
phase transition. Two
examples can be condensed matter under extreme external conditions or
certain stages of
the quark-gluon plasma formation process. \par

In section 2 we formulate the problem and obtain an integral expansion in 
the parameter $m_0/E$, where $m_0$ is the height of the wall. We 
consider separately analytic and non-analytic functions as wall profiles. 
Section 3 is devoted to the study of functions defined as zero outside
 a certain region, in which the wall is located, where the
potential varies.
These functions must be non-analytic on the real axis as will be seen. In 
section 4 general analytic functions are treated. An imaginary, chirally odd
and CP violating mass 
term is incorporated into the formalism in section 5.
Finally, in section 6 we summarize and conclude.  

\bigskip

\section{The integral expansion for the Dirac equation solutions}

\bigskip

As usual in this kind of problems, we work in the rest frame of the
wall, which is taken
to be planar and parallel to the $x$-$y$ plane. Thus, the mass will be
characterized by a $z$-
dependent function, the wall profile function. The approximation for a
planar interface for the bubble wall should be valid for large bubbles
compared to the
microscopic size scale. In the electroweak phase transition this is
valid for most of
the evolution of bubbles \cite{[9]}. For simplicity we study
the Dirac equation
for one flavour. In order to calculate the reflection coefficient we
need only the plane
wave solution for particles moving along the $z$-axis. In any case,
general solutions for
other incoming directions can be obtained by performing the appropriate
Lorentz boost in
the $x$-$y$ plane. \par

Following Nelson et al. \cite{[6]} \cite{[10]}, we work in the chiral
basis and factor
the Dirac operator into $2 \times 2$ blocks. Thus the Dirac equation
can be expressed as
\beq
\pmatrix{
i \partial_z + i \partial_t &- m^*(z) &0 & 0 \cr
m(z) &i \partial_z - i \partial_t &0 &0 \cr
0 &0 &i\partial_z + i \partial_t &- m(z) \cr
0 &0 &m^*(z) &i \partial_z - i \partial_t \cr
} \Psi = 0 \label{1} 
\eeq

\noindent and using the following ansatz for solutions with positive energy $E$
\beq
\Psi =  {\psi_I \choose \psi_{II}} e^{-iEt}
 \qquad \hbox{with} \quad \matrix{\psi_I = {\psi_1 \choose \psi_2}  \cr
\psi_{II} = {\psi_3 \choose \psi_4} \cr } {\label{2}}
\eeq
\noindent we obtain
\bea
&&\left ( i \partial_z + Q(z) \right ) \psi_I = 0 \nn \\
&&\left ( i \partial_z + \overline{Q}(z) \right ) \psi_{II} = 0 \label{3} 
\eea

\noindent where
\beq
Q(z) = \pmatrix{ E &-m^*(z) \cr
m(z) &- E \cr
} \qquad , \quad \overline{Q}(z) = \pmatrix{E &-m(z) \cr
m^*(z) &-E \label{4} \cr
} \ .
\eeq
\noindent Taking into account that the eigenvectors of the chiral basis have been
reordered to obtain $\gamma^5 = \pmatrix{\sigma_3 &0 \cr 0 &- \sigma_3 \cr}$, the
eigenvalues of this matrix, the chirality, will be + 1 for $\psi_1$
and $\psi_4$, and -
1 for $\psi_2$ and $\psi_3$. \par The solutions of the equation
(\ref{3}) can be written
as follows 
\bea
&&\psi_I(z) = {\cal P} e^{i \int_{z_0}^z d\tau Q(\tau )} \psi_I(z_0) \
\ \ , \nn \\   
&&\psi_{II}(z) = {\cal P} e^{i \int_{z_0}^z d \tau \overline{Q}(\tau )}
\psi_{II} (z_0) \
\ \ . \label{5}
\eea
Where ${\cal P}$ indicates a path ordered product and $\tau$ is the position 
variable along the $z$-axis. The Dirac equation solutions are determined by
\bea
&&\Omega (z, z_0) = {\cal P} e^{i \int_{z_0}^z d \tau Q(\tau )} \qquad
\hbox{and} \nn \\
&&\overline{\Omega}(z, z_0) = {\cal P} e^{i \int_{z_0}^z d \tau
\overline{Q}( \tau )} \ \
\ , \label{6} 
\eea
\noindent which, in  a matrix way, can be written as
\beq
\Omega (z, z_0) = \pmatrix{\omega_{11} &\omega_{12} \cr
\omega_{21} &\omega_{22} \cr
} \qquad , \quad \overline{\Omega} (z , z_0) = \pmatrix{ \overline{\omega}_{11}
&\overline{\omega}_{12} \cr \overline{\omega}_{21} &\overline{\omega}_{22} \cr
} \ \ \  .  \label{7}
\eeq
\noindent It is very easy to show that $\omega_{21} = \omega_{12}^*$
and $\omega_{22} =
\omega_{11}^*$ and analogously for $\overline{\Omega}(z, z_0)$. Thus,
it follows from
(\ref{5}) that, knowing $\omega_{11}$, $\omega_{12}$, $\overline{\omega}_{11}$ 
and
$\overline{\omega}_{12}$, we have the solutions of the time-independent Dirac 
equation . \par

If we consider $m(\tau ) = 0$ for $\tau < z_0$, then $\psi_1$ and $\psi_2$
correspond to right-moving right-handed particles and left-moving left-handed
particles respectively, as well as $\psi_3$ and $\psi_4$ to
right-moving left-handed and
left-moving right-handed particles, in this region. If $m(\tau )$ is
considered to be a
certain constant, $m$, for $\tau > z$, $\Omega (\tau , z)$ and
$\overline{\Omega}(\tau
,z)$ can be immediately diagonalized in order to identify the right-moving and the
left-moving flux of particles. Thus, by requiring that the left-moving
flux is zero for
$\tau > z$, i.e. allowing only a transmitted flux beyond the wall,
 a left-handed reflected flux is
obtained from the right-handed incident one and vice versa. Therefore
\beq
{\cal R}_{R \to L} = |R(E)|^2 \qquad \hbox{and} \qquad {\cal R}_{L \to R} = |
\overline{R}(E)|^2 \label{8} 
\eeq
\noindent where ${\cal R}_{R \to L}$ represents the probability for a
reflected left-handed flux from an incident hight-handed one and ${\cal
R}_{L \to R}$ that for a reflected right-handed flux from an incident
left-handed one. From the CPT theorem 
we have that ${\cal R}_{\bar{L} \to \bar{R}} = {\cal R}_{R\to L}$ and 
${\cal R}_{\bar{R} \to \bar{L}} = {\cal R}_{L \to R}$, where 
$\bar{R}$ and $\bar{L}$ label left and right antifermions, respectively\cite{[6]}, 
\cite{[5]}. We obtain for the reflection coefficients
\bea
&&R(E) = - e^{2iEz_0} {\omega_{12}^* - {E - P \over m_0} \omega_{11}
\over \omega_{11}^*
- {E - P \over m_0} \omega_{12}} \ \ \ , \nn \\
&&\overline{R}(E) = - e^{2iEz_0} {\overline{\omega}_{12}^* - {E - P \over m_0}
\overline{\omega}_{11} \over \overline{\omega}_{11}^* - {E - P \over m_0}
\overline{\omega}_{12}} \ \ \ ; \label{9} 
\eea
\noindent with $p = + \sqrt{E^2 - m_0^2}$. A more detailed derivation of 
(\ref{9}) may be found in refs. \cite{[6]} and \cite{[10]}$^1$. Now the
task is to evaluate $\Omega (z, z_0)$ and $\overline{\Omega}(z, z_0)$. We
consider at first the real mass case whence from
(\ref{4}) that $Q(\tau ) =
\overline{Q}(\tau )$. Thus, $\Omega(z, z_0) = \overline{\Omega}(z, z_0)$ and 
consequently ${\cal R}_{R \to L} = {\cal R}_{L \to R}$. 
Therefore, it is obvious that if the mass is real in (\ref{4}) no CP-violating 
effect exists. \par

In this case $Q(\tau )$ can be expressed as:
$Q(\tau ) = \sigma_3E - i \sigma_2 \ m_0 \ f(\tau )$,  
with $m(\tau ) = m_0 f(\tau )$ and ${\vec \sigma}$ are the Pauli 
matrices. We require $f(\infty ) = 1$ and
$f(\tau ) = O[1]$. 
In what follows the characteristic energy and length, $m_0$ and 
${1 \over m_0}$, are used to get dimensionless quantities, so that eq.
(\ref{10}) must be written as:

\beq
Q^*(\tau ) = \sigma_3 E^* - i \sigma_2 \ f(\tau ) \ \ \ , \label{10}
\eeq
\noindent

where $Q^*(\tau ) = {Q(\tau ) \over m_0}$ and $E^* = {E \over
m_0}$. We do not mark the dimensionless quantities
with an asterisk in what follows. \par

As shown in appendix A, taking into account the anti-commutation
properties of the
Pauli matrices and the definition of $\Omega (z, z_0)$ as a path ordered product,
the latter can be expanded as follows
\beq
\Omega (z, z_0) = \left ( \sum_{n=0} \Gamma_n(E) \left ( {1 \over E} \right )^n
\right ) e^{i\sigma_3 \int_{z_0}^z d \tau \ p(\tau )} \label{11} 
\eeq
\noindent where 
\beq
p(z) = +(E^2 - [f(\tau )]^2)^{1/2} \ \ \ ,  \label{12}
\eeq 
\noindent and the first terms of the
expansion can be written as
\bea
&&\Gamma_0 = 1 \ \ \ , \nn \\
&&\Gamma_1(E) = \sigma_2 \int_{z_0}^z d \tau \  p(\tau ) \ f(\tau ) \ e^{-2i\sigma_3
\int_{\tau}^z d\xi p(\xi)} \ \ \ , \nn \\ 
&&\Gamma_2(E) = \int_{z_0}^z d\tau \int_{z_0}^{\tau} d\xi \ p(\tau ) \ p(\xi ) \
f(\tau ) \ f(\xi ) \ e^{-2i\sigma_3 \int_{\xi}^{\tau} d \chi p(\chi )} + \cr 
&&\hskip 1.5 truecm + i {\sigma_3
\over 2} \int_{z_0}^z d\tau \ p(\tau ) (f(\tau ))^2 \ \ \ , \nn \\
&&\Gamma_3(E) = \nn \\
&&\sigma_2 \Big \{ \int_{z_0}^z d\tau \int_{z_0}^{\tau} d\xi
\int_{z_0}^{\xi} d\chi \ p(\tau ) \ p(\xi ) \ p(\chi ) \ f(\tau ) \
f(\xi ) \ f(\chi ) 
e^{-2i\sigma_3 \int_{\tau}^z d \rho p(\rho )} e^{-2i \sigma_3
\int_{\chi}^{\xi} d\rho 
p(\rho )} \nn \\ 
&&+ {i \sigma_3 \over 2} \int_{z_0}^z d \tau \int_{z_0}^{\tau} d \xi
\left [ f(\tau )
(f(\xi ))^2 e^{-2i \sigma_3 \int_{\tau}^z d\chi p(\chi )} - f(\xi )
(f(\tau ))^2 e^{-2i
\sigma_3 \int_{\xi}^z d\chi p(\chi )} \right ] p(\tau ) p(\xi ) + \nn \\
&&+ {1 \over 2} \int_{z_0}^z d \tau \ p(\tau ) \ (f(\tau ))^3 \ e^{-2i \sigma_3
\int_{\tau }^z d\chi p(\chi )}  \Big \} \ \ \ . \label{13}
\eea  
In principle, the expansion (\ref{11}) must be convergent as a consequence of the
definition of $\Omega (z, z_0)$. Nevertheless, an explicit proof is not
easy. In any case, it can be directly shown \cite{[11]},
 although tediously, that (\ref{11}) is a good asymptotic 
expansion by using a diagrammatic notation and certain prescriptions to obtain the
mentioned expansion in a general way.
This asymptotic character is enough for our high energy considerations. \par

In next sections we will study separately  non-analytical profile
functions for
a wall with finite real thickness and analytical profile functions for
a wall extending formally to infinity,
although with a finite effective thickness because the mass goes exponentially 
to zero when $\tau \to - \infty$ and to $m_0$ when $\tau \to +\infty$. 

\bigskip

\section{The non-analytical case}

\bigskip

We consider that the profile functions describes a finite domain wall
which extends from $\tau = - \delta_w$ to $\tau = \delta_w$ (obviously 
$\delta_w$ is the half of the total wall thickness). In general, a certain function
in the region $(- \delta_w$, $\delta_w$) is matched onto $f(\tau ) = 0,1$ for 
$\tau < -\delta_w$ and $\tau > \delta_w$, respectively. There is no analytical 
function able to describe the wall structure from $z_0 < - \delta_w$ 
to $z > \delta_w$ because at least a derivative of a certain order does not 
exist at the points $\tau = -\delta_w,\delta_w$. All the derivatives
defined to the left of $\tau = -\delta_w$ or to the right of $\tau = \delta_w$ 
are zero, 
\beq
\lim_{\tau\to -\delta_w^-} f^{(n)}(\tau ) = 
\lim_{\tau \to \delta_w^+} f^{(n)}(\tau ) = 0 \quad , \qquad \forall \
n > 0 \ \ \ ; \label{14} 
\eeq 

\noindent nevertheless, there must be certain $k_0$, $k_1$ that
verify 
\beq
\lim_{\tau \to -\delta_w^+} f^{(k_0)}(\tau ) \not= 0 \quad , 
\quad \lim_{\tau \to \delta_w^-}
f^{(k_1)} (\tau ) \not= 0 \ \ \ . \label{15}
\eeq
\noindent The $k_0$-derivative and the $k_1$-derivative are not continuous
functions on $\tau$, they are not well-defined for $\tau = -\delta_w,\delta_w$, 
respectively and $f(\tau )$ is not analytical on the real axis. \par

In order to study this case it is convenient to introduce the new
integration variable
$x = {\tau \over \delta_w}$ and the parameter $a = 2E\delta_w$. We consider the 
evolution inside the
domain wall (i.e. from $z_0 = - \delta_w$ to  $z = \delta_w$). In principle
we take $a$ and $E$ as two different parameters in the expansion. Following these
prescriptions and expanding $p(\tau )$ in powers of $\left ( 1/E \right )$ we
can derive from (\ref{13})
\bea
&&\Gamma_1(E) = \sigma_2 e^{-i\sigma_3a} \Big \{ {a \over 2} \int_{-1}^1 dx \ 
F(x) \ e^{i\sigma_3ax} - \left ( {1 \over E} \right )^2 {a \over 4} 
\int_{-1}^1 dx \ (F(x))^3 \ e^{i \sigma_3ax} \nn \\
&&\hskip 1.5 truecm + i \sigma_3 \left ( {1 \over E} \right )^2 {a^2 \over 4} 
\int_{-1}^1dx \int_x^1 dy (F(y))^2 F(x) \ e^{i\sigma_3ax} \Big \} \nn \\
&&\Gamma_2(E) = {a^2 \over 4} \int_{-1}^1 dx \int_{-1}^x dy \ F(x) \ F(y) \ 
e^{-i\sigma_3ax} \ e^{i\sigma_3ay} + {i\sigma_3 \over 4} a \int_{-1}^1 dx 
(F(x))^2 \nn \\
&&\Gamma_3(E) = \sigma_2 \ e^{-i\sigma_3a} \Big \{ {a^3 \over 8} \int_{-1}^1 dx
\int_{-1}^x dy \int_{-1}^y dt \ F(x) \ F(y) \ F(t) \ e^{i\sigma_3 a(x - y + t)} 
\nn \\
&&\hskip 1.5 truecm + {i \sigma_3 \over 8} a^2 \int_{-1}^1 dx \int_{-1}^xdy 
\left [ F(x)(F(y))^2 \ e^{i\sigma_3ax} - F(y) (F(x))^2 \ e^{i\sigma_3ay} 
\right ] \nn \\
&&\hskip 1.5 truecm + {a \over 4} \int_{-1}^1 dx (F(x))^3  \ e^{i\sigma_3ax} 
\Big \} \ \ \ .
\label{16}  
\eea
\noindent For convenience we define $F(x) = f(\delta_w x)$. By assuming that 
$E^2 \gg a$$^2$, the exponentials containing integrals of $F(x)$ 
have been expanded 
in $\left ( {a \over E^2} \right )$. In (\ref{16}) the expansion is considered 
up to terms $O\left (\left ( 1/E \right )^3 \right )$. 
In general, the comparison between (\ref{7}) and
(\ref{11}) gives expressions for $\omega_{11}$ and $\omega_{12}$ depending on 
the functions $\Gamma_n(E)$, the former is written as an expansion for $n$ 
even and the latter for $n$ odd. Follows from (\ref{16}) and after some 
calculations  up to order
$O\left [ \left ( 1/E \right )^3 \right ]$ 
\bea 
&&\omega_{11} = \left [ 1 + {1 \over 4}
\left ( {1 \over E} \right )^2 \left ( R_{2,2}(a) + i R_{2,1}(a) \right ) + o\left [
\left ( {1 \over E} \right )^3 \right ] \right ] \ e^{i\phi}  \\ &&\omega_{12} = - i
e^{ia} \left [ {1 \over 2} \left ( {1 \over E} \right ) R_{1,1}^*(a) +
{1 \over 8} \left
( {1 \over E} \right )^3 \left ( R_{3,3}^*(a) - iR_{1,1}^*(a)
R_{2,1}(a) \right ) + o
\left [ \left ( {1 \over E} \right )^3 \right ] \right ] e^{-i\phi} \ , \nn
\label{17}
\eea
\noindent where
\bea
&&R_{1,1}(a) = a \int_{-1}^1 dx \ F(x) \ e^{iax} \nn \\
&&R_{2,1}(a) = a \int_{-1}^1 dx (F(x))^2 \nn \\
&&R_{2,2}(a) = a^2 \int_{-1}^1 dx \int_{-1}^x dy \ F(x) \ F(y) \
e^{ia(y - x)} \nn \\
&&R_{3,3}(a) = a^3 \int_{-1}^1 dx \int_{-1}^x dy \int_{-1}^y dt \ F(x)
\ F(y) \ F(t) \
e^{ia(t - y + x)} \ \ \ . \label{18} 
\eea
\noindent $\phi$ is defined as the classical path, $\phi= \int_{z_0}^z d\tau
\ p(\tau )$, and
for convenience the expansion of the exponentials of $\phi$ will be considered 
only at the end. \par

{From} $\omega_{11}$ and $\omega_{12}$ we can obtain, eq. (\ref{9}),  a general
expansion for the reflection coefficient $R(E)$. It can be seen that
the numerator of
(\ref{9}) only contains odd powers of ${1 \over E}$, while the even ones only 
are in the denominator. As a consequence of this fact, by expanding the 
denominator, $R(E)$ can be written as a series of only odd powers in 
${1 \over E}$. It is convenient to write
\beq
R(E) = e^{2i(\phi - a)} R_0(E) \ \ \ , \label{19} 
\eeq
\noindent where
\beq
R_0(E) = \sum_{n=1} C_n(a) \left ( {1 \over E} \right )^n \ \ \ . \label{20}
\eeq
\noindent It is interesting to stress that the phase factor $e^{2i(\phi - a)}$ which
multiplies $R_0(E)$ depends on the difference between the classical path and the
classical free path neglecting the mass function. This phase will be of
the order $O[({1 \over
E})^2]$. \par

{From} eq.(\ref{17}), the first coefficients for (\ref{20}) can be
identified. We have 
\bea
&&C_1(a) = {1 \over 2} \left ( e^{ia} - i R_{1,1}(a) \right ) \nn \\
&&C_3(a) = {1 \over 8} \left [ e^{ia} + i R'_{3,3}(a) + \left ( e^{ia}
- i R_{1,1}(a)
\right ) \left ( 2 R_{2,1}(a) - \right . \right . \nn \\
&&\hskip 1.5 truecm \left . \left . - e^{ia} R_{1,1}^*(a) + 2i R_{2,2}^*(a) - i
|R_{1,1}(a)|^2 \right ) \right ] \ \ \ , \label{21}  
\eea
\noindent where the functions $R_{1,1}(a)$, $R_{2,2}(a)$ and
$R_{2,1}(a)$ are defined in
(\ref{18}). In order to express the final result in a more
compact way, $R_{33}$ can be written in terms of $R'_{33}$ plus another term which
will be canceled during the calculation. $R'_{33}$ defined as follows

\beq
R'_{3,3}(a) = a^3 \int_{-1}^1 dx \int_x^1 dy \int_{-1}^y dt \ F(x) \ F(y) \ F(t) \
e^{ia(t - y + x)} \ \ \ . \label{22} 
\eeq
\noindent The validity of (\ref{21}) is general provided that the
profile function is
strictly equal to 0 for $\tau < - \delta_w$ and $1$ for $\tau > \delta_w$. 
We will now check this result by considering a simple example as the step
profile function, $f(\tau ) = \theta (\tau )$. Then $F(x) = \theta (x)$
from $x = - 1$ to
$x = 1$. Obviously the parameter $\delta_w$ lacks any physical meaning
and we hope that
the final result is independent on $a$. In fact, the exact reflection
coefficient has
been directly calculated for this case (see for example
\cite{[5]}) and we have
\beq
R(E) = {1 \over p + E} \ \ \  , \label{23}
\eeq
\noindent where $p$ has been defined
in (\ref{12}). By expanding (\ref{23}) in powers of ${1 \over E}$, we obtain 
\beq
R(E) = {1 \over 2}\left ( {1 \over E} \right ) + {1 \over 8}
\left ( {1 \over E} \right )^3 + o\left [ \left ( {1 \over E} \right
)^3 \right ] \ \ \
. \label{24} \eeq 
\noindent On the other hand, from (\ref{18}), (\ref{21}) and (\ref{22})
applied with the
step function $F(x) = \theta (x)$ we obtain
\bea
&&C_1(a) = 1/2 \ \ \ , \nn \\
&&C_3(a) = {1 \over 8} (1 + 2ia) \ \ \ . \label{25} 
\eea
\noindent Therefore, taking into account that $\phi = a - {a \over 4}
\left ( {1 \over
E} \right )^2 + o\left [ \left ( {1 \over E} \right )^3 \right ]$ in this case, it
follows from (\ref{19}) and (\ref{20}) 
\bea
&&R(E) = e^{-i {1 \over 2}\left ( {1 \over E} \right )^2a + o \left [
\left ( {1 \over
E} \right )^3 \right ]} \left \{ {1 \over 2} \left ( {1 \over E} \right
) + {1 \over 8}
\left ( {1 \over E} \right )^3 (1 + 2ia) + o\left [ \left ( {1 \over E}
\right )^3 \right
] \right \} = \nn \\
&&\hskip 1.2 truecm = {1 \over 2} \left ( {1 \over E} \right ) + {1
\over 8} \left ( {1
\over E} \right )^3 + o \left [ \left ( {1 \over E} \right )^3 \right ] \ \ \ .
\label{26} \eea
\noindent Thus, we have a positive check for (\ref{21}). \par

Now, we study the problem in a general way. By considering the
arguments exposed in the
beginning of this section, we introduce a certain function $F(x)$
between $x = - 1$ and
$x = 1$, in such a way that the function and all its $k$-derivatives up
to $k = h - 1$ are
matched onto $F(x) = 0, 1$ for $x < - 1$ and $x > 1$, respectively.
Therefore, $F(x)$
must verify
\bea
&&F^{(k)}(1) = F^{(k)}(-1) = 0 \quad , \qquad 0 < k < h \ \ \ ; \nn \\
&&F(1) = 1 \ \ \ , \nn \\
&&F(-1) = 0 \ \ \ . \label{27}
\eea
\noindent With these requirements it follows from (\ref{18}) and (\ref{21}) that
\beq
C_1(a) = \sum_{k=h}^{\infty} {i^k \over a^k} \ A_k \ e^{i \varphi_k} \label{28} 
\eeq
\noindent where$^3$

\bea
&&A_k = {1 \over 2} \left [ \left ( F^{(k)}(1) \right )^2 + \left (
F^{(k)}(-1) \right
)^2 - 2 \cos 2a \ F^{(k)}(1) \ F^{(k)}(-1) \right ]^{1/2} \ \ \ , \nn \\
&&\varphi_k = \pi + {\arctan} \  \left [ {F^{(k)}(1) + F^{(k)}(-1)
\over F^{(k)}(1) -
F^{(k)}(-1)} {\tan} \ a \right ] \ \ \ . \label{29}
\eea

\noindent To obtain (\ref{28}), the following result has been used

\beq
\int^x d\tau \ F(\tau) e^{ia\tau} = {e^{iax} \over ia} \sum_{k=0} \left
( {i \over a}
\right )^k \ F^{(k)}(x) \ \ \ . \label{30}
\eeq
\noindent Last eq. (\ref{30}) can be usually found for any function
$F(x)$ which can be written as a finite power series\cite{[12]}.
Nevertheless, it is possible to prove it to be valid more 
generally for any power series provided that the r.h.s. of eq. (\ref{30}) is absolutely 
convergent\cite{[11]}. Moreover, by successively integrating by parts, each one of
the different orders in $1/a$ appears and the asymptotic character of
such series, which is the only requirement we need,  can be proven provided that $F(x)$ 
is not a quasi-periodic
function (i.e. it cannot be written as the product of a positive
function times a periodic one). We will therefore concentrate 
on non quasi-periodic profiles, although an interesting physical
consequence of the quasi-periodic ones, which can be well understood in
our formalism, must be shown before. The decreasing with the energy
of the reflection coefficient and the resulting CP asymmetry 
(as will be seen in section 5) is due to the position dependent optical phase,
$e^{iax}$, in our formulas. A larger energy produces oscillations of a
minor period for the integrand in the left hand-side of eq. (\ref{30}),
in such a way that the integral goes to zero, decreasing as energy
increases. Nevertheless, if the profile function is quasi-periodic
positive interferences can take place. In fact, when the energy is
taken so that both periods, the one for the oscillations due to the
optical phase and the one of the quasi-periodic profile, are the same,
the integrand is defined positive and a peak is expected around
this value. Moreover, by taking into account that $a=2 E \delta_w$,
this peak shifts towards larger energies when either a thinner wall is
considered or the period of the profile decreases. For the latter case,
a minor period will also generate a concentration of the asymmetry
around the peak, since the effect of the oscillations due to the
profile outside the energy range of the peak will be the same as those
of the high energy optical phase. These effects were pointed out in
ref. \cite{Tor96} for a particular quasi-periodic profile wall. 

Until now, $E$ and $a$ have been considered as different parameters in the
expansion. In fact, we expand in ${1 \over E}$ for large $E$ and
consider $a$ fixed. In
this case (\ref{28}) gives the leading term in the asymptotic behaviour of the
reflection coefficient. The convergence of the sum depends in general on the
convergence of the right-hand side of (\ref{30}), and the good
asymptotic behaviour of the latter guarantees that of the former. Nevertheless, $a$
fixed for $E$ large
implies $\delta_w$ small and in this way the profile function
considered is representing
a thin wall, thinner for larger $E$. By comparing the reflection
coefficient obtained from
(\ref{20}) and (\ref{28}) with (\ref{26}), we can see that a power law for 
the fall with the energy is obtained in both cases, and in the limit $a
\to 0$ the result 
for the step case is recovered. It is easy to show from (\ref{18}) 
and (\ref{21}) that 
\bea
&&C_1(a) = {1 \over 2} + O[a] \ \ \ , \nn \\
&&C_3(a) = {1 \over 8} + O[a] \ \ \ . \label{31} 
\eea
\noindent It results that the validity of the step profile functions
demands, beyond the
obvious condition $\delta_w \ll 1$, the additional one $a \ll 1$. With the
appropriate dimensions, these conditions are 
$\delta_w \ll \hbar/m_0c$ and $E \ll \hbar c/\delta_w$, respectively. The latter giving
the range of energy where the fermions are not sensitive to the real functional form of 
the wall. \par

A more interesting situation is when $\delta_w$ is kept fixed and 
$a$ increases with
$E$. The expansion (\ref{20}) must be rewritten in terms of $E$ and $\delta_w$. We
obtain
\beq
R_0(E) = \sum_{k=h} {\cal D}_k(\delta_w , E) \left ( {1 \over E} \right
)^{k+1} \ \ \
, \label{32} 
\eeq
\noindent where
\beq
{\cal D}_h(\delta_w, E) = {i^h \over (2 \delta_w )^h} A_h \ e^{i \varphi_h} \ \ \ .
\label{33} \eeq
\noindent The remaining dependence on $E$ of the coefficients ${\cal D}_k$ in 
(\ref{32}) is due to the factors $e^{\pm ia}$ which appear in the integrals of 
(\ref{18}) and they are
not expandable in powers of ${1 \over E}$. It is worth to stress that the
exponential factor $e^{2i(\phi - a)}$ in (\ref{19}) can be expanded in powers 
of $\left ( 1/E \right )$ 
because, in this term, the phase goes to 0 as $E \to \infty$. This
energy dependence of ${\cal D}_k$ does not affect the good asymptotic
properties of the
expansion (\ref{32}), i.e. ${\cal D}_h\left ( {1 \over E} \right
)^{h+1}$ gives the high
energy behaviour of the reflection coefficient, since the factors $e^{\pm ia}$, and
therefore the coefficients ${\cal D}_k$, remain bounded when $E \to
\infty$. These factors $e^{\pm ia}$
are crucial for the study of the CP-violating process, as will be seen in 
section 5, because they provide a non-trivial CP-even phase.
Eqs.~(\ref{32}) and (\ref{33}) are derived from
(\ref{20}) and (\ref{28}) provided that the expansion in eq. (\ref{20}) keeps 
its asymptotic character as the dependence on $E$ of the coefficients $C_n(a)$ 
through the parameter $a$ is considered.
The latter happens when  

\beq
\lim_{E \to \infty} \ {\left ( {1 \over E} \right )^{k+2} \left |
C_{k+2}(2 \delta_w E
) \right | \over \left ( {1 \over E} \right )^k \left | C_k(2 \delta_w
E) \right | }= 0
\label{34} 
\eeq
 
\noindent for all odd $k$. The observance of this condition can be proven 
although it is not an immediate consequence of the
asymptotic character of the expansion in (\ref{11}) because the profile
function $f(\tau)$ is not zero outside the domain wall~: $f(\tau ) = 1$ for 
$\tau \geq \delta_w$. It
follows from (\ref{21}) that (\ref{34}) is satisfied for $k = 1$. A
general proof may be
found in \cite{[11]}. \par

Therefore, for the functions introduced in this section as mathematical
representation
of a strictly finite physical wall, we obtain a fall with the energy as
the power $- (h
+ 1)$, where $h$ is the order of the first derivative which is not
continuous (or does
not exist) at any of the two point $\tau = 0, 1$, or at both. This result can 
be easily generalized in order to assert that the {\it dominant 
 high 
energy behaviour, when the  profile function is $h$ times derivable on the real axis, its $h$-derivative, 
being discontinuous, is $\sim 1/E^{h+1}$ as in
eq.~(\ref{32})}. This is the main result of this section.

\bigskip

\section{Analytic profile functions}

\bigskip
 
We consider now the case where the profile function is analytic on the real axis. By
considering the arguments we presented in the previous section it
follows immediately
that an analytic function cannot describe a wall profile which is
constant outside the
domain wall. Therefore, these profile functions can only be
asymptotically equal to 0,
1, as $\tau \to - \infty$, $+ \infty$, respectively. Thus, the wall extends
formally from $\tau = - \infty$ to $\tau = + \infty$. Nevertheless,
this wall will be
characterized in general by an effective thickness, $\sigma$, which
allows to define a
certain domain wall. The particular criterion considered in order to
define $\sigma$ is
not important. For instance, $|f(\tau ) - 1| < 0.1$ for
$\tau > \sigma$ and $|f(\tau )| < 0.1$ for $\tau < - \sigma$. However,
in general we can
write $f(\tau ) = F\left ({\tau \over \sigma} \right )$, where we will assume
\beq
\mathrel{\mathop {F(x)}_{x \ll - 1}} \ \sim e^{x C_-} \qquad , \qquad
\mathrel{\mathop
{F(x)}_{x \gg + 1}}\ \sim 1 - e^{-xC_+} \ \ \ ; \label{35}
\eeq
 \noindent being $C_+ > 0$, $C_- > 0$, parameters depending on the
particular profile wall. The
asymptotic functional behaviour assumed in eq.~(\ref{35}) is in
principle the simplest
that can be demanded to a profile function verifying the initial
asymptotic conditions
as $\tau \to - \infty$, $+ \infty$ introduced above. Furthermore, it is not
easy to find functional behaviour other than that of (\ref{35}), with good analytic
properties on the real axis, and, on the other hand, a naive derivation of the
wall profile from the corresponding equation of motion as done in
\cite{[7]} agrees with
this assumption$^4$.  
We consider the evolution from $\tau = - \delta_w$ to $\tau = \delta_w$,
but this parameter is now defined in such a way that $f(\delta_w x) = F(\lambda
x)$, where $\lambda = {\delta_w \over \sigma}$ verifies $1 \ll \lambda \ll {E \over
\sigma}$. Thus, analogously to the previous section, it follows from (\ref{11})
and (\ref{13})  \beq
\Omega (\lambda \sigma,-\lambda \sigma) = \left \{ 1 + {1 \over E}
\sigma_2 {a \over 2} \int_{-1}^1 dx \ F(\lambda x) \ e^{-i \sigma_3 a(1 - x)} 
+ o \left [ \left ( {1 \over E} \right )^2 \right ] \right\}
e^{i\sigma_3 \phi} \ \ \ ; \label{36}
\eeq
\noindent where all the previous definitions are kept. \par

Again, we define $R(E) = e^{2i(\phi - a)} R_0(E)$ and obtain
\beq
R_0(E) = {1 \over 2} \left ( {1 \over E} \right ) \left ( e^{ia} - i a
\int_{-1}^1 dx
\ F(\lambda x) e^{iax} \right ) + o\left [ \left ( {1 \over E} \right
)^2 \right ] \ \ \
. \label{37}
\eeq
\noindent In order to derive (\ref{37}) from (\ref{36}), the first of the equations
(\ref{9}) must be used, and therefore the profile function is modified
outside the region
$(- \delta_w , \delta_w )$ to describe a situation with zero and
constant mass for $\tau
< - \delta_w$, $\tau > \delta_w$. This technical assumption, necessary to
identify the incident, reflected and transmitted particle fluxes outside the region
considered, introduces non-negligible effects for the high energy
behaviour which must be
separated from the real ones caused by the evolution through the wall region
considered. In fact, the numerical calculation performed in \cite{[6]} 
requires this technical assumption and as a consequence the authors consider a 
wall strictly located in the region between $z = 0$ and $z = z_0$ despite they 
work with an analytical ansatz which goes exponentially to zero and one when 
$z \to - \infty$, $+\infty$, respectively. However, this modification has no 
practical consequence except for the high energy behaviour. It should be 
noted that the hypothesis $E \gg \delta_w$ is necessary to obtain (\ref{36}), 
therefore we cannot take the limit $\delta_w \to \infty$ directly in order to 
calculate the reflection coefficient. When the profile function varies from 
$\tau = - \infty$ to $\tau = + \infty$, the coefficients must be formally 
defined by using the asymptotic flux at $\tau = + \infty$, $-\infty$. 
Obviously, it should be done before obtaining eq. (\ref{36}),
by considering the evolution of the fermions from $\tau = - \infty$ to $\tau
= + \infty$ , without the introduction of the parameter $\delta_w$ and without any
additional condition for the energy beyond the initial one, $E \gg 1$. Nevertheless,
in this case we cannot expand the exponentials of (\ref{13}) to derive an analogous
expression to (\ref{16}), the integrals we obtained for the expansion of the
reflection coefficient are then much more complex and practically it is
not possible to
analyze them in a general way. This is why we approach the problem in a
different way:
we study in practice the
evolution from $-\delta_w$ to $\delta_w$ in order to find the total contribution
from $- \infty$ to $+ \infty$ and other terms depending on $\delta_w$,
which must be in
principle negligible by considering the requirement (\ref{35}) for the profile
function.\par

By rewriting appropriately the integral in (\ref{37}) as the sum of two
terms, the first
giving the contribution of the whole integration domain and the second due to
contributions of $x \notin [-1, 1]$,  we obtain after some calculation,
for the leading
term, 
\beq
R_0^{(1)}(E) = - i \left ( {1 \over E} \right ) {a \over 2} \left \{
\lim_{\varepsilon
\to 0^+} \int_{- \infty}^{+ \infty} dx \ F(\lambda x) \ e^{(ia -
\varepsilon)x} \right \}
+ \left ( {1 \over E} \right ) R_{N-C}(a) \label{38} 
\eeq
\noindent where, from (\ref{35}) and $\lambda \gg 1$ 
\beq
R_{N-C}(a) = + {i \over 2} \left \{ e^{-\lambda C_+} \ e^{ia} {a \over
ia - \lambda C_+} + e^{-\lambda C_-} \ e^{-ia} {a \over ia + \lambda
C_-} \right \} \ \ \
. \label{39}
\eeq
\noindent It can be seen that the first term of the right-hand side of
(\ref{38}) does not depend on ${\delta_w}$ (Although it seems to have a
dependence on $\delta_w$ through $a$, if we change the integration variable $x$ by
$x' = \lambda x$, we see that this dependence disappears). 
Now, we analyze its functional dependence on the energy and then we
turn to the question whether $R_{N-C}(a)$ is negligible with regard to
this term. \par

If the Laurent expansion for $F(z)$, the analytic complex extension of $F(x)$, 
in the pole of order $\nu_j$, $z = z_j$, is
$\sum\limits_{n=-\nu}^{+ \infty} a_n^j (z - z_j)^n$ the following
result can be shown by
applying the Cauchy theorem
\beq
\lim_{\varepsilon \to 0^+} \int_{-\infty}^{+\infty} dx \ F(\lambda x) \ e^{(ia -
\varepsilon )x} = 2 \pi i \sum_{j=1}^N e^{- {a \over \lambda} y_j} \ e^{i{a \over
\lambda} x_j} \sum_{n = 1}^{\nu_j} a_{-n}^j {(ia)^{n-1} \over
\lambda^n(n - 1) !} \ \ \
. \label{40}
\eeq
\noindent Where $z_1 = x_1 + iy_1$, $z_2 = x_2 + i y_2$, ..., $z_N =
x_N + iy_N$, are
all the poles of $F(z)$ with positive imaginary part we have picked
when the integration
contour is adequately closed. As the profile function, $F(x)$, is
analytic on the real
axis, it is obvious that $y_j \not= 0$ and therefore we obtain an
exponential dependence
on $a$. Nevertheless, in this case the parameter $a$ has no physical
 meaning because
$\delta_w$ is arbitrary. The distance $\delta_w$ is taken as several
times the parameter $\sigma$ in order to apply eq. (\ref{35}).
Moreover, $\sigma$ is defined after an arbitrary criterion. The
right-hand side of eq. (\ref{40}) may however be rewritten as follows

\beq
{2 \pi i \over \lambda} \sum_{j=1}^N e^{-2E \sigma y_j} \
e^{2iE\sigma x_j} \sum_{n=1}^{\nu_j} a_{-n}^j {(2iE \sigma )^{n-1}
\over (n - 1) !} \ \ \
, \label{41}
\eeq 

\noindent where the dependence on $\delta_w$, through the parameter
$a$, is replaced by the one on $\sigma$. The final result will be
expressed, (47),  in terms of the singularities of $f(z)$, instead $F(z)$, 
eliminating 
any dependence on our arbitrary parameter $\sigma$ 
\cite{Rod96}.

Now, the discussion of (\ref{41}) is very similar to the previous one
for the non-analytic case. If the range of energy and the effective
thickness, $\sigma$,
allows to consider that the factor $E\sigma$ is not large, then the
term ${1 \over E}
R_{N-C}$ is negligible with regard to the first one of eq. (\ref{38}), 
\beq
\sigma E \ e^{-2E\sigma y_j} \gg e^{-\lambda C_+} \quad , \quad e^{-\lambda C_-} \ \
\  ; \label{42} 
\eeq
  
\noindent since $\lambda$ is large$^5$ the latter gives then the asymptotic behaviour of the
reflection coefficient.
Nevertheless, by considering the energy dependence of (\ref{41}) and
$R_{N-C}(a)$, it is
obvious that for large enough energy the first term of (\ref{38}) is
smaller than ${1
\over E} R_{N-C}$ since (\ref{42}) is not satisfied. This result is at
first surprising and
seems to disagree with our requirements about the profile function. But, now
we must consider the above mentioned technical assumption since as a
consequence of that,
the profile function we are studying in practice is non-continuous at
$\tau = - \delta_w$
and $\tau = \delta_w$. Therefore, a contribution of the order ${1 \over
E}$ to some power follows from
the results of the previous section which obviously must depend on
$\delta_w$. It is not easy to obtain exactly these contributions because
 the result
(\ref{30}) cannot be in general applied to these profile functions,
except in the large $a$ limit
 where the asymptotic behaviour can be determined. It can be 
proven that the asymptotic behaviour for large $E$, and therefore $a$, of the 
integral in the left-hand side of eq. (\ref{30}) is indeed given by the first term 
in the right-hand side of eq. (\ref{30}), in a completely general way for any 
function $F(x)$. Thus, the contribution of the non-continuities can be 
identified. Moreover, if we take the limit $\sigma \to 0$ then $R_{N-C}
\to 0$, which is the behaviour we hope for the contribution of
non-continuities: the profile function gives the step function, in this
limit, and it can also be proven that the contribution of the
right-hand side's first term of eq. (\ref{38}) is exactly the
reflection coefficient for the step case \cite{Rod96b}.
The energy dependence of the term ${1 \over E} R_{N-C}$ is 
explained in this way, as well its surprising 
large energy dominance. However we
are interested in the high energy contribution of the total evolution from $\tau = -
\infty$ to $\tau = + \infty$, and all the previous discussions allow to
assume that it is
given by the first term of eq. (\ref{38}). By using eq. (\ref{41}) we
obtain for this high energy
behaviour of the reflection coefficient 
\beq
R(E) = 2 \pi \sigma \sum_{j=1}^N e^{-2E\sigma y_j} \
e^{2iE\sigma x_j} \sum_{n=1}^{\nu_j} a_{-n}^j {(2iE \sigma )^{n-1}
\over (n - 1) !} \ \ \
. \label{43}
\eeq

Besides, if we are in the range of energy where $E \sigma $ is large, the leading
contribution is given by the pole of $F(z)$ with the smallest imaginary
part, i.e. the
closest to the real axis. Calling $z_k = x_k + i y_k$ this prevailing pole we obtain
\beq
R(E) = 2 \pi \sigma e^{-2E\sigma y_k} \ a_{-\nu_k}^k \ {(2i E \sigma )^{\nu_k - 1}
\over ( \nu_k - 1) !} \ e^{2iE\sigma x_k} \ \ \ . \label{44}
\eeq

\noindent In this case there is no proof of a result analogous to
(\ref{34}) (even the
particular checks are difficult). However, by studying directly (\ref{13}) as 
well as the prescription to obtain the coefficients of the general expansion 
(\ref{11}) which is given in ref. \cite{[11]}, it can be
reasonably guessed that the asymptotic character of eq. (\ref{20}) is given by 

\beq
\lim_{E \to \infty} {\left ( {1 \over E} \right )^{k+2} \left |
C_{k+2}(2 \sigma E ) \right | \over \left ( {1 \over E} \right )^k
\left | C_k(2 \sigma
E) \right |} \ \ \lsim \sigma^2 \ \ \ . \label{45}
\eeq

\noindent  Looking at (\ref{13}) and taking into account the expansion
of eq. (\ref{12}) for $p(\tau)$ in powers of $(1/E)$, it is easy to see
that $\Gamma_3(E)$ contains terms of the order 
$E^3 \sigma^3$, because the first of the three integrals of
$\Gamma_3(E)$ in eq. (\ref{16}) appears multiplied by $(\sigma E)^3$
when the change of integration variable, $x={\tau \over \sigma}$, 
is performed after the introduction of the function $F(\tau) = f(\sigma \tau)$.
If we conjecture that this
result can be generalized to $\Gamma_n(E)$ containing terms of the
order $E^n \sigma^n$ (this conjecture is supported by the study of the general expansion
given by eq. (\ref{11}) in ref. \cite{[11]}), we get the ratio given by eq. (\ref{45}).
If the particular profile function considered generates an expansion
for the reflection
coefficient which verifies (\ref{45}), then (\ref{43}) and (\ref{44})
will be valid in any
range of energy satisfying $E \gg 1$ and 
$\sigma \ll 1$ (obviously (\ref{44}) is only
valid for $\sigma E$ large). As can be seen above, these conditions are
suitable for studying quarks propagating through a wall except for the
case of the top. \par

We worked by convenience with the function 
$F(z)$, obtained through the change of variables performed at the beginning 
of the present section. Let us now return to the physical variables and 
express the final result as a function of the singularities of $f(z)$ instead of $F(z)$ 
in order to avoid the apparent arbitrariness arising from the dependence on $\sigma$.
If we consider the Laurent expansion 
$\sum\limits_{n=-\nu_j}^{+ \infty} a_n^{'j} (z - z'_j)^n$, for $f(z)$ in the 
pole of the order $\nu_j$, $z=z'_j$  and take into account that 
$f(z)=F({z\over \sigma})$, we obtain

\bea
&& z'_j = \sigma z_j \ , \nn \\
&& a^{'j}_n = {a_n\over \sigma^n} \ .
\label{ZandB}
\eea

\noindent
Eq. (\ref{44}) can thus be re-written as follows

\beq
R(E) = 2 \pi e^{-2E y'_k} 
\ a_{-\nu_k}^{'k} \ 
{(2i E)^{\nu_k - 1}\over ( \nu_k - 1) !} \ 
e^{2iE x'_k} \ ;
\label{FinEq}
\eeq

\noindent where the former condition $\sigma E \gg 1$  
must be re-formulated by requiring that $e^{2E(y'_k-y'_h)}$ may be neglected, 
$y'_k$ ($y'_h$) being the imaginary parts of the (next to) closest pole to the 
real axis. It is also interesting to notice that the wall thickness 
we introduced through the parameter $\sigma$ is now incorporated without 
arbitrariness into the profile function.
From eq. (\ref{ZandB}), the connection between this thickness and the distance between 
the poles of $f(z)$ is apparent.

Now, we check the present results by using the particular analytic solution 
obtained in references \cite{[7]} and \cite{[8]} for the ansatz
\beq
f(\tau ) = {1 + {\rm \tanh} \ (\tau / \sigma ) \over 2} \ \ \ . \label{46}
\eeq

For this profile function it follows from (\ref{43}) that
\beq
R(E) = {\pi \sigma \over 2 \sinh (\pi E \sigma )} \ \ \ , \label{47}
\eeq
\noindent which agrees with the result of \cite{[7]} and  \cite{[8]} in
the above mentioned
conditions. \par

Therefore, with this positive check, we conclude that (\ref{43}) in general and
(\ref{44}) for $E \sigma$ large give the high energy behaviour of the reflection
coefficient, for the analytic profile functions: 
{\it at high energy the reflection coefficient decreases exponentially, the coefficient in the exponential being the smallest positive imaginary part of the complexified profile function}. This is the main result of this section.

It is interesting to emphasize the
exponential dependence on the energy in this case instead the power
dependence in the
previous one. Apparently, there is no connection between 
 power
 law energy dependence derived in in the preceding section
for non-analytic profiles, and the exponential law 
for analytic profiles derived in the present section. Nevertheless
it can be proven that the results for non-analytic profiles can be
understood through the properties in the complex plane of analytic
extensions for these profile functions by using eq. (\ref{43}) \cite{Rod96b}. 
\par

\bigskip

\section{CP-violating wall profiles}

\bigskip

As pointed out in the introduction, our effort
 to study CP-violating effects concentrates on CP violation at the
tree level due to the expectation value of the Higgs fields beyond the
Standard Model. 
The time independent Dirac Hamiltonian incorporating CP-violation can
be written as follows

\beq
H= {\vec \alpha } {\vec p} + \beta m_R(z) +i \beta \gamma^5 m_I(z) \ .
\label{CPHam}
\eeq

\noindent As stated in ref. \cite{[5]}, $m_R(z)$ and $m_I(z)$
are real because of the hermiticity of the Hamiltonian. The CP violation 
comes from the term
$\propto m_I(z)$, when it cannot be rotated away  through a chiral rotation.
Therefore, we must go beyond the SM in order to find CP violation at
tree level. In the SM, with one Higgs field $m_I(z)/m_R(z)$ is a constant 
independent of $z$ and therefore this source of CP violation can be rotated 
away. In the other models, with more than one Higgs field, 
$m_I(z)/m_R(z)$ depends in general on $z$ and  the
matrix $m_I(z)$ will remain.

Following the notations and conventions of section 2.  the time
independent Dirac equation can be written as in 
eqs. (\ref{3}) and (\ref{4}), where $m(z)= m_R(z)+im_I(z)$. The
CP-violating effects are so incorporated by assigning a complex mass to
the fermion inside the domain wall. 

Analogously to (\ref{10}), we define $m_R(\tau) = m_0 f(\tau )$ and
$m_I(z)= \varepsilon m_0 g(\tau)$, where $f(\tau )$ and $g(\tau )$ are
real functions of order 1
which characterize the CP-violating profile, and $\varepsilon$ is a
parameter associated to CP-violation. We express
\bea
&&Q(\tau ) = \sigma_3 E - i \left ( \sigma_2 f(\tau) - \varepsilon
\sigma_1 g(\tau) \right ) \
\ \ , \nn \\
&&\overline{Q}(\tau ) = \sigma_3 E - i \left ( \sigma_2 f(\tau) +
\varepsilon \sigma_1 g(\tau)
\right ) \ \ \ . \label{48}
\eea
\noindent We keep the same dimensionless quantities, the same definitions of the
previous sections and assume that $g(\tau)$ goes exponentially to zero at
both, $\tau \to +
\infty$, $\tau \to - \infty$, for an analytic profile or that it it is strictly
zero outside the
domain wall for a non-analytic one. This choice is always possible by an 
appropriate chiral rotation which rotates away the imaginary part $m_I(+\infty)$
, i.e. which corresponds to a real mass in the broken phase. We have checked 
that the same result is obtained with another choice.

 Thus, the high energy {\it leading} term for the
reflection coefficient can be expressed in a general way as
\beq
\left \{ \matrix{R^{(1)}(E) \cr \overline{R}^{(1)}(E) \cr} \right \} = {1 \over 2}
\left ( {1 \over E} \right ) \left ( e^{ia} - ia \int_{-1}^1 dx \
F(\lambda x) \ e^{iax}
\pm a \varepsilon \int_{-1}^1 dx \ G(\lambda x) \ e^{iax} \right ) \ \
\ , \label{49}
\eeq
\noindent where $G(x)$ is defined as $F(x)$ in the previous section,
$x$ being the same dimensionless variable which was then introduced,
and $\lambda = 1$, $\delta_w/\sigma$ for non-analytic, analytic
profiles, respectively. The sign $-$ is for $R(E)$ and $+$ for
$\overline{R}(E)$. \par

For both, analytic and non-analytic profiles, the difference ${\cal R}_{R \to
L}$-${\cal R}_{L \to R}$, which is a quantitative measure of the CP-violating 
effects induced by the wall, can be computed. For the sake of simplicity we 
concentrate on the non-analytic ones:

\beq
{\cal R}_{R \to L}-{\cal R}_{L\to R} = -4 \varepsilon \left ( {1 \over E} \right
)^{h+\ell + 2} {A_h \ B_{\ell} \over (2 \delta_w)^{h+\ell}} 
\sin (\Delta \varphi ) \ \ \
, \label{50}
\eeq 
\noindent where the $\ell$-derivative of $G(x)$ has been considered as the
first non-continuous one, where $B_{\ell}$ is defined from $G(x)$ as $A_h$ in
(\ref{29}) from $F(x)$,  and where
\beq
\Delta \varphi = \alpha_{\ell} - \varphi_h + (\ell - h) {\pi \over 2} \
\ \ , \label{51}
\eeq
 \noindent with  $\varphi_h$ defined in (\ref{29}) from $F(x)$ 
 and $\alpha_{\ell}$ analogously from
$G(x)$. When $g(\tau)$ vanishes, $B_{\ell}=0$ and as expected the r.h.s of (\ref{50}) vanishes. \par

The former requirement, $g(\tau)$ vanishing outside the domain wall, is
imposed as above mentioned in order to work in the physical basis, in which the  
masses are real. For technical reasons,  we will also consider the chirally rotated convention in which
$g(\tau)= 1$ in the broken phase outside the domain wall, i.e.
$m_I=\varepsilon m_0$. 

One can easily check from (\ref{48}) that the imaginary mass in the broken 
phase will be removed by the chiral rotation 
$\exp (i\eta\gamma_5)$ with $\varepsilon=\tan(2\eta)$, and the real mass 
in this physical region will become $m= m_0 |\sqrt{1+\varepsilon^2}|$.
As a matter of the fact, in that region the eigenvalues of the matrix
$Q(z)$ and $\overline{Q}(z)$ will be $\pm p$, where
$p=|\sqrt{E^2-m^2}|$, $m$ obviously being the physical mass. Therefore,
the dimensionless quantities we introduced in previous sections should
be now built by using $m$ instead $m_0$. Then, eq. (\ref{49}) must be replaced 
by

\bea
\left \{ \matrix{R^{(1)}(E) \cr \overline{R}^{(1)}(E) \cr} \right \} = {1 \over 2}
\left ( {1 \over E} \right ) \left ( e^{ia}{1\pm i\varepsilon \over
\sqrt{1+\varepsilon^2}} - {ia \over \sqrt{1+\varepsilon^2}} \int_{-1}^1 dx \
F(\lambda x) \ e^{iax} \right. \\ \nonumber
\left. \pm {a \varepsilon \over \sqrt{1+\varepsilon^2}} \int_{-1}^1 dx
\ G(\lambda x) \ e^{iax} \right ) \ \ \ , \label{51B}
\eea

\noindent and it follows that

\beq
{\cal R}_{R \to L}-{\cal R}_{L\to R} = -{4 \varepsilon \over 1+\varepsilon^2} 
\left ( {1 \over E} \right
)^{h+\ell + 2} {A_h \ B_{\ell} \over (2 \delta_w)^{h+\ell}} 
\sin (\Delta \varphi ) \ \ \
, \label{51C}
\eeq 

\noindent instead of eq. (\ref{50}). By comparing eqs. (\ref{50}) and
(\ref{51C}) it is very easy to see that all physical consequences 
and in particular the discussion which follows
will be valid for both cases. For example, the vanishing of the r.h.s. of  
(\ref{51C}) when no CP violation is present is now due 
to $\Delta \varphi=0$.
An amusing remark is that 
eq. (\ref{51C}) shows that the highest chiral asymmetry originated by
the CP-violating wall profile appears for the case $\varepsilon = 1$.
In other words, by taking into account that as well $f(\tau)$ as
$g(\tau)$ are required to be $O[1]$, the most effective situation in what 
concerns the production of a net baryon number left in the broken phase for 
the high energy range arises then from requiring that real and pure 
imaginary mass terms are of the same order. In fact, Eq. (\ref{51C}) remains unchanged 
if we make $\varepsilon \to {1\over \varepsilon}$, this symmetric behaviour for 
$\varepsilon > 1$ and $\varepsilon < 1$ being coherent with the fact that 
real and pure imaginary masses in the broken phase can be turned into each
other through a chiral rotation. As already stated, the imaginary mass
in the broken phase can be rotated away by a chiral rotation to
return from the latter convention, $g(\infty)=1$, to the former one in which
$g(\tau)$ vanishes outside the domain wall. The chiral 
rotation implies a redefinition of the new real and pure imaginary functions, 
${\cal F}(\tau)$ and ${\cal G}(\tau)$, as a function of the old ones,
$f(\tau)$ and $g(\tau)$. If we then apply the new functions, ${\cal
F}(\tau)$ and ${\cal G}(\tau)$, which behave now appropriately outside
the domain wall, to eq. (\ref{49}) and obtain the first non-null
contribution to the chiral asymmetry, eq. (\ref{51C}) will be
re-obtained. This expected result is not immediate to obtain explicitly, 
because the
contribution to the chiral symmetry given by eq.(\ref{50}), which is in
general non-null, gives zero for ${\cal F}(\tau)$ and ${\cal G}(\tau)$
if $h \neq \ell$. Therefore upper order terms must be considered to
re-obtain eq. (\ref{51C}).

As well known, a CP-violating effect is always generated by the interference
between a CP-even phase and a CP-odd one. In fact, it can be seen from
eqs. (\ref{49}), (\ref{50}) and (\ref{51}) that a non-zero leading term
first needs the CP-odd phase arising from the reversing of the sign for
the pure imaginary mass term in (\ref{49}), but also a non-zero angle
$\Delta \varphi$, i.e. a net CP-even phase. Notice  that working at high energy
($E\gg 1$) we are far off the total reflection domain ($E\le 1$), and that due to the thickness of the wall nothing prevents a CP-even phase. In fact the CP-even phase is generated by the interference of the phase of the mass term $m_I(z)/m_R(z)$
at different values of $z$. 
In the step approximation,
it is always possible to make the mass
term real by rotating away the phase through a chiral rotation, but if we have a
$z$-dependent complex mass term inside a certain region, this rotation
is not possible
in general and the imaginary term generally remains. Tree level effects  can
thus appear. Nevertheless, if the $z$-dependence of the imaginary term
is proportional
to the real one, we can, again through a chiral rotation, turn the complex mass into real. When we take the convention $g(+\infty)=0$ this situation corresponds to $g(z)=0$ for all z.
This is stressed in references \cite{[6]} and \cite{[8]} and it
is obvious from (\ref{29}) that in
this case $h = \ell$, $\varphi_h = \alpha_{\ell}$ and therefore $\Delta
\varphi = 0$,
which lead to ${\cal R}_{R\to L}$-${\cal R}_{L \to R} = 0$.
Furthermore, the leading term
will be zero if we assume a weaker condition than the proportionality
of real and pure
imaginary terms. \par

Assuming $F(x) = C + F_0(x)$ and $G(x) = K + G_0(x)$, where $C$, $K$ are simple
constants and $F_0(-x) = - F_0(x)$, $G_0(-x) = - G_0(-x)$, we obtain
\bea
&&F^{(h)}(1) = (-1)^{h+1} \ F^{(h)}(-1) \ \ \ , \nn \\
&&G^{(\ell)}(1) = (-1)^{\ell + 1} \ F^{(\ell )}(- 1) \ \ \ ; \label{52}
\eea
\noindent for $h$, $\ell > 0$. It follows from (\ref{52}) and
(\ref{51}) that under these
conditions \beq
\Delta \varphi = n \pi \ \ \ , \label{53}
\eeq
\noindent where $n$ is an integer. Analogously we obtain (\ref{53}) if
both functions have
even symmetry with regard to the axis $x = 0$ (It is obvious that our
requirement $F(+ \infty ) = 1$ forbids an even symmetry for $F(x)$, but our results can be applied to
general situations where $F(+ \infty ) = 0$). We conclude therefore that this weaker
condition implies a null CP-violating leading term as it follows from
(\ref{50}). This does not lead to a strict suppression of CP-violating
effects as the stronger
requirement of proportionality does, but it can be stated that
CP-violation will be of a lower order in the high energy range. 
Whether these symmetry conditions involve any similar consequence in a
lower energy range is an interesting question, but it is out of the
scope of the present
work. \par

It is also interesting to stress that the most favourable situation is given by the
opposite symmetry properties of $F(x)$ and $G(x)$ with regard to the
axis $x=0$. If we assume that the functions $F_0(x)$ and $G_0(x)$,
above defined, are even and odd symmetric, respectively, or vice versa,
we immediately obtain
\beq
\Delta \varphi = {2n + 1 \over 2} \pi \ \ \ , \label{54}
\eeq
\noindent $n$ being an integer. The highest value for $\sin (\Delta
\varphi )$ is thus given by imposing these symmetry requirements.\par

An expression analogous to (\ref{50}) can be found for the analytic case by using
(\ref{44}), and a similar discussion may be done. The crucial
point for such a discussion is that the integral
$\int_{-\infty}^{\infty}dx F(\lambda x) e^{(i a - \varepsilon)x}$,
which is obtained from $\int_{-1}^1 dx F(\lambda x) e^{i a x}$, is pure
imaginary if $F(x)$ differs by any constant from an odd symmetric
function, and real if it is even symmetric.
It is very easy to see that, for instance, the analytic ansatz
(\ref{46}) is odd symmetric
with regard to the axis $x=0$ up to a  constant, its
contribution to the reflection coefficient has indeed no CP even 
phase factor (eq. (\ref{47})). If a function $F(x)$ for the real mass term 
 and a function $G(x)$  for the imaginary one have both odd symmetry,
  the angle $\Delta
\varphi$ vanishes. No CP-even phase is obtained and no
CP-violating effects at tree-level and at first order in the high
energy limit arise from this case. Nevertheless, if a function $G(x)$
defined as ${dF(x) \over dx}$, which is even symmetric, is taken, a net
CP-even phase and therefore CP-violating effects at the first order are
obtained. This simple ansatz for $G(x)$ illustrates our
statement about the symmetry properties. 

In ref. \cite{[8]} numerical calculations are performed by considering
the CP-violating term as a perturbation to the CP conserved Dirac
equation, for the particular real profile function above mentioned,
$F(x)=1/2(1+ \tanh x)$, and for the two following pure imaginary ones

\bea
G_1(x)={dF(x) \over dx} \ \ ; \nonumber \\
G_2(x)=\left( F(x) \right)^2 \ \ . 
\label{N59}
\eea 

\noindent 
The quantity $\Delta^{CP}={\cal R}_{R \to L}-{\cal R}_{L \to R}$ for
both particular cases can be immediately obtained in the high energy
range from the general results given by eqs. (\ref{44}) and (\ref{49}),

\bea
\Delta^{CP}_1 = 8 \pi^2 \varepsilon \sigma^3 E e^{-2\pi E
\sigma} \ \ ,  \nonumber \\
\Delta^{CP}_2 = - 4 \pi^2 \varepsilon \sigma^3 E e^{-2\pi E
\sigma} \ \ .  \label{N60}
\eea

\noindent These results are valid obviously for $E \gg 1$ and $\sigma
\ll 1$ as we explained in section 4, but the requirement $E \sigma \gg
1$ is also necessary to isolate the contribution of the leading
singularity \cite{Rod96b}. Nevertheless, the contribution for all the
poles can be resummed and we obtain, for instance, in the first of the
two former cases

\bea
\Delta^{CP}_1 = 2 \varepsilon \sigma^3 E
\left( {\pi \over \sinh(\pi \sigma E)} \right)^2 \ \ , \label{N61}
\eea

\noindent which is a valid result for any $E \sigma$.

The energy range we are interested in is not explored in the numerical
results given in ref. \cite{[8]}, although a few results for $E=5.0$
are presented amusingly in relatively good agreement with the ones 
we obtain from our
analytic expression$^6$. We also checked in
section 4 that the unperturbed reflection coefficient which is obtained
in that work for the above expressed particular ansatz completely
agrees in the appropriate limit with our general result for a non
CP-violating wall profile. Furthermore, we obtain the ratio
$\Delta^{CP}_1/\Delta^{CP}_2=-2$ which is observed for the results
given by Funakubo {\it et al.} \cite{[8]}, even for $E \sim 1$.

Several interesting remarks must be finally done. The function
$G_2(x)=[F(x)]^2$ does not present (up to a constant) any well defined symmetry with respect 
 to the axis $x=0$, while $G_1[x]={dF(x)
\over dx}$ is even symmetric; a higher CP asymmetry for the latter than
for the former is thus not surprising in view of our conclusions about symmetry.
Moreover, the dependence on $\sigma$ of eq. (\ref{N61}) also agrees
with the well-known fact that no observable CP-asymmetry can appear in
the thin wall case at tree level, i.e. in the limit $\sigma \to 0$.
Funakubo {\it et al} state that the degree of decreasing of
$\Delta^{CP}$ seems to be larger as the wall thickness increases after
studying two particular cases for several values of the wall thickness. 
Regarding eq. (\ref{50}) we can state a more precise conclusion for any
non-analytic wall profile in the high energy range which confirms the
trend of the CP asymmetry to decrease when thickness increases pointed
out for those authors. The quantity $\Delta^{CP}$ generally is an
oscillating function on the thickness parameter because the angle
$\Delta \varphi$ depends on it, but those oscillations are modulated by
a factor decreasing with larger thickness as eq. (\ref{50}) shows
($\Delta^{CP}$ as a function of the thickness has in general an
infinite number of zeros, the first of them being in fact as the
thickness is zero). The functional behaviour of $\Delta^{CP}$ given in
eqs. (\ref{N60}) and (\ref{N61}) for analytic wall profiles 
seems also confirm that trend, but in that case our results are only
valid for $\sigma \ll 1$.

\bigskip

\section{Summary and conclusions}

\bigskip
 
Rigorous and completely general conclusions about the behaviour of
fermions hitting a
wall with an arbitrary profile are very difficult. Obviously, the
fermion scattering is
determined by the particular profile function which characterizes the wall we are
treating. The crucial point in order to get general answers in this
problem is to show
relations between the functional properties of the wall profile and the
reflection coefficients, etc, which characterize the behaviour of the
fermions hitting
the wall. A prescription to solve the Dirac equation in the presence of CP-violating
electroweak bubble wall is presented in \cite{[8]}. Nevertheless, the results 
depend on certain functions of the energy which must be obtained solving a
second-order differential equation. This prescription, applied to the
particular ansatz
studied in \cite{[7]}, leads to a hypergeometric differential equation,
but in general it
is a difficult problem which does not allow to relate in a direct and
useful way the wall
profile function to the fermion behaviour. \par

A formalism to study the scattering of the high energy fermions in the presence of a
general wall is proposed in the present work. High energy fermions mean
$E/m_0 \gg 1$. If
we consider that the temperature during the supposedly electroweak first order phase
transition is about 100 GeV \cite{[14]}, the Boltzmann thermal
distribution gives an average energy for the quarks which verifies ${m_0 \over E} =
O[10^{-2}]$ except for the top. Thus, the range of energy considered is
interesting in
the cosmological problem, although for the top we only study a high
energy tail of the
thermal distribution. The quantum corrections to the classical
behaviour expected for
this range have been obtained for two different types of mathematical
representations of
the wall. The first describes walls with profile functions which differ from the
asymptotic values only in a finite domain. Whether they are realistic is a difficult
question. They are modeling situations involving different length
scales of variation
for the profile function and its derivatives, i.e. so that to a given accuracy, some
derivatives can be considered as discontinuous. In the second case, when we
assume that the wall profile function differs from the asymptotic values up to
infinity, edge effects related with different length scales of
variation are neglected
(quantum, thermal and statistics fluctuation are not considered, for
example). However,
we study the quantum corrections in both cases and show how the high
energy fermions are
sensitive to these facts. As can be expected, the quantum corrections
are more important
as the profile becomes sharper. In fact, if the $h$-th derivative of the
profile function is the first  non
continuous one, i.e. it presents a certain step, the high energy behaviour
of the reflection coefficient is characterized
by the following power law 

\beq
R(E) \propto m_0 c^2(\hbar c/\delta_w)^h (1/E)^{h+1} \ \ , 
\label{propN-A}
\eeq

\noindent against the exponential behaviour 

\beq
R(E) \propto \exp[-y'_k E /(\hbar c)] \ \ ,
\label{propA}
\eeq

\noindent given by a profile
function analytic on the real axis. The physical dimensions are restored in 
eqs. (\ref{propN-A}) and (\ref{propA}). $\delta_w$ in the first equation is the 
 wall thickness parameter defined in section 3 and the coefficient $y'_k$ 
in the latter one is the smallest positive imaginary part
among the poles of the analytic complex extension of the profile function 
$f(\tau)$. It worth to stress that $y'_k$ has the dimension of a length, 
since the variable $\tau$ for $f(\tau)$ characterizes the position in the 
normal axis to the planar wall. 
Notice also that, while in the former case the reflection 
coefficient is proportional to some power in $\hbar$ as expected for
a quantum effect, in the latter case the 
reflection coefficient is not even analytic in $\hbar$, similar to a tunnelling 
effect (for example, the  term given by 
eq. (\ref{propA}) has an analogous dependence on $\hbar$
 to that of the so-called {\it Gamow} factor, $e^{-2\pi Z Z'e^2/\hbar v}$,
which gives the dominant behaviour for processes involving nuclei with charge $Z'e$ and
velocity $v$ tunneling through the Coulomb barrier due to a charge 
$Ze$\cite{Schi68}).

Finally, CP-violating effects have been
incorporated by introducing a complex mass, as we explained in section 5. 
The reflection coefficients, $R(E)$ and $\overline{R}(E)$, are expressed
as a function of both, real, chirally-even, and pure imaginary, chirally odd
mass functions, $f(z)$ and $g(z)$. We first perform a chiral rotation to 
have a real mass for $z\to \infty$, deeply in the broken phase. 

If $g(z)$ does not identically vanish, the phase of the mass depends on the chirality of the incoming flux. The resulting chiral asymmetry
 leads to a CP-asymmetry through the CPT
theorem. This CP-violating term introduced
in the Dirac Hamiltonian produces a CP-odd phase. This CP-odd phase is not enough to generate an
observable CP asymmetry. It is known that a CP-even phase should exist
and interfere with the CP-odd one. That effect is clearly shown in our
formalism through the dependence of the relevant quantity $\Delta^{CP}
= {\cal R}_{R \to L} - {\cal R}_{L \to R}$ on the angle $\Delta
\varphi$. A non-zero angle is due to the CP-even phases arising from
the integration of the real and pure imaginary terms of the profile
function, although the asymptotic regime considered here is far off the total reflection. We found that the respective symmetry properties of the $f(z)$ and $g(z)$ play an important role, and we obtain the
following general conclusions: i) CP-even phases and therefore
CP-violating observable effects at tree level will be found 
in the thick wall case, provided that
 the symmetry properties
 are not the same for both, $f(z)$ and $g(z)$. ii) CP-violating effects will be of lower 
order in $1/E$ (if not vanishing) when the symmetry properties are the
same for both functions. iii) the most favourable
situation in order to generate CP-asymmetry in this high energy range
is given by a profile where real and pure imaginary functions present
opposite symmetries.

It is finally very important to remark that the axis $z=0$ is an
arbitrary reference for the symmetry properties of functions.
As expected our results are obviously independent of any shift on
$z$, the physical conclusions thus depending on the respective symmetry
properties of the real and imaginary mass terms. It is also important to notice that the conclusion about symmetry properties are valid even if the symmetry properties appear after a chiral rotation which in general leads to a non vanishing $g(+\infty)$.

\bigskip

\section*{Acknowledgments.}
We are specially indebted to Jean-Claude Raynal for early inspiring 
discussions and in particular for mentioning that the singularities 
of the wall profile functions might play an important role. We also 
acknowledge M. Calvet for typing the first version of this work.

\bigskip

\appendix

\section{The integral expansion}

\vskip 5 truemm

In this appendix we derive the integral expansion given in equations
(\ref{11}) and (\ref{13}). 

The starting point is the definition given in (\ref{6}) for $\Omega (z, z_0)$
which leads to the solutions of the time-independent Dirac equation as shown 
in (\ref{5}). We had

\beq
\Omega (z, z_0) = {\cal P} \ e^{i \int_{z_0}^z d\tau \ Q(\tau )} \ \ \ , 
\eeq

\noindent where $Q(\tau )$ is expressed by means of the usual Pauli's matrices in
(\ref{10}). By considering a certain path partition $(z_0, z_1, ...,
z_{N-1}, z_N, z)$,
we can write

\beq
{\cal P} \ e^{i\int_{z_0}^z d\tau \ Q(\tau )} = {\cal P} \ e^{i \int_{z_N}^zd \tau \
Q(\tau )} {\cal P} \ e^{i\int_{z_{N-1}}^{z_N} d\tau \ Q(\tau )} ... \
{\cal P} \ e^{i
\int_{z_0}^{z_1} d\tau \ Q(\tau )} \ \ \ . \label{A2}
\eeq

\noindent If we take $Q(\tau )$ as constant in each interval $(z_j ,
z_{j+1})$, defining
for each one $f_j$ as the following integral average value

\beq
f_j = {1 \over \Delta_j} \int_{z_j}^{z_{j+1}} d \tau \ f(\tau ) \ \ \ , 
\eeq

\noindent where $\Delta_j = z_{j+1} - z_j$, it can be written by 
using (\ref{10})

\beq
{\cal P} \ e^{i\int_{z_j}^{z_{j+1}} d\tau \ Q(\tau )} = e^{(i\sigma_3E + \sigma_2
f_j)\Delta_j} \ \ \  . \label{A4}
\eeq

\noindent Taking into account the following result for two operators verifying 
$\{A,B\} = 0$ and $A^2 = B^2 = 1$ 

\beq
e^{\alpha A + \beta B} = \cosh \left [ \left ( \alpha^2 + \beta^2
\right )^{1/2} \right
] + {\alpha A + \beta B \over \left ( \alpha^2 + \beta^2 \right
)^{1/2}} \sinh \left [
\left ( \alpha^2 + \beta^2 \right )^{1/2} \right ] \ \ \ , \label{A5}
\eeq

\noindent which can be easily proven, it follows from (\ref{A4}) that:

\bea
&&{\cal P} \ e^{i \int_{z_j}^{z_{j+1}} d \tau \ Q(\tau )} = e^{i\sigma_3 p_j
\Delta_j} + \nn \\
&&+ \left \{ {1 \over E} \sigma_2 f_j + \left ( {1 \over E} \right )^2 {i \over 2}
\sigma_3 f_j^2 + \left ( {1 \over E} \right )^3 {\sigma_2 \over 2} f_j^3 + o\left [
\left ( {1 \over E} \right )^3 \right ] \right \} \sin (p_j \Delta_j) \ \ . 
\label{A6} 
\eea 

\noindent where $p_j = + (E^2 - f_j^2)^{1/2}$. In order to obtain (\ref{A6}) 
${E \over p_j}$ and ${1 \over p_j}$ have been expanded in powers of 
${1 \over E}$. Eq. (\ref{A5}) is also valid
for any set of unitary operators anticommuting with each other and 
consequently analogous result to the following can be obtained for the complex mass 
case. \par

By substituting the path ordered products of the right-hand in (\ref{A2}) by 
(\ref{A6}), multiplying and reordering the terms, we obtain

\bea
\Omega (z, z_0) = e^{i\sigma_3 \sum\limits_{j=0}^N p_j \Delta_j}   
+ {1 \over E} \sigma_2 \sum\limits_{k=0}^N \sin (p_k \Delta_k)f_k \ 
e^{-i\sigma_3\sum\limits_{j=k+1}^N p_j \Delta_j} \ 
e^{i\sigma_3 \sum\limits_{j=0}^{k-1} p_j \Delta_j} \nn \\ 
+ \left ( {1 \over E} \right )^2 \left \{ \sum\limits_{k=0}^N 
\sum\limits_{m=0}^{k-1}
\sin (p_k \Delta_k ) \sin (p_M \Delta_m) f_k \ f_m \ 
e^{i\sigma_3 \sum\limits_{j=k+1}^N
p_j \Delta_j} \ e^{-i \sigma_3 \sum\limits_{j=m+1}^{k-1} p_j \Delta_j}  
e^{i \sigma_3 \sum\limits_{j=0}^{m-1} p_j \Delta_j} \right . \nn \\
\left .  
+ i {\sigma_3 \over 2}
\sum\limits_{k=0}^N \sin (p_k \Delta_k) f_k^2 \ 
e^{i\sigma_3 \sum\limits_{j=0}^N p_j\Delta_j} \right \} \nn \\ 
+ \left ( {1 \over E} \right )^3 \sigma_2 \left \{
\sum\limits_{k=0}^N \sum\limits_{m=0}^{k-1} \sum\limits_{n=0}^{m-1} 
\sin (p_k \Delta_k)
\sin (p_m \Delta_m) \sin (p_n \Delta_n) f_k \ f_m \ f_n \right . \nn \\ 
\times \ e^{-i \sigma_3\sum\limits_{j=k+1}^N p_j \Delta_j} \ 
e^{i\sigma_3 \sum\limits_{j=m+1}^{k-1} p_j\Delta_j} \ 
e^{-i \sigma_3 \sum\limits_{j=n+1}^{m-1} p_j \Delta_j} \ e^{i\sigma_3
\sum\limits_{j=0}^{n-1} p_j \Delta_j} \nn \\
+ {i \sigma_3 \over 2}
\sum\limits_{k=0}^N \sum\limits_{m=0}^{k-1} \sin (p_k \Delta_k) \sin
(p_m \Delta_m) f_k \
f_m^2 \ e^{-i \sigma_3 \sum\limits_{j=k+1}^N p_j \Delta_j} \ e^{i\sigma_3
\sum\limits_{j=0}^{k-1} p_j \Delta_j} \nn \\ 
- {i \sigma_3 \over 2} \sum\limits_{k=0}^N
\sum\limits_{m=0}^{k-1} \sin (p_k \Delta_k) \sin (p_m \Delta_m) f_k^2 f_m \ e^{-i
\sigma_3 \sum\limits_{j=m+1}^N p_j \Delta_j} \ e^{i\sigma_3
\sum\limits_{j=0}^{m-1} p_j
\Delta_j} \nn \\ 
\left . + {1 \over 2} \sum\limits_{k=0}^N \sin (p_k \Delta_k )f_k^3 \
e^{-i\sigma_3 \sum\limits_{j=k+1}^N p_j \Delta_j} \ e^{i\sigma_3
\sum\limits_{j=0}^{k-1}
p_j \Delta_j} \right \} + ... \  , \label{A8}
\eea
\noindent where the sum in 
the exponential is taken to be zero if the lower index is bigger than the 
upper. The general result for two operators $A$, $B$, which
verify $\{A, B\} = 0$, $e^{B}A = Ae^{-B}$ has been used. 
We know that $\sin (p_j \Delta_j)= p_j \Delta_j
+ O[\Delta_j^2]$ and in the limit $\Delta_j \to 0$, 
$\sum\limits_j\Delta_j \to \int d\tau$ with $f_j \to f(\tau )$ and 
$p_j \to p(\tau )$. These replacements can be 
understood if we consider the definition of $f_j$ as an integral average value,
assuming that $f_j = f(\tau_j )$ with $\tau_j \in [z_j , z_{j+1}]$, provided that
$f(\tau )$ is continuous. Following the last prescriptions, Eq.
(\ref{11}) and (\ref{13})
can be immediately derived from (\ref{A8}). 

\bigskip

\section{General Orthogonality properties}

\bigskip

In the present appendix we consider in a general way the problem of the 
orthogonality of different Dirac Hamiltonian eigenstates in the presence of a 
wall. We keep the same framework presented in section 2, assuming a z-dependent 
mass in order to study the one flavour quark propagation problem. Let 
$(E,p_x,p_y,p_z)$ be the four-moment of the quark. For a static wall the 
particle energy, $E$, is conserved. In the unbroken phase $E=p_z$. The 
system is symmetric with respect to rotations around the z-axis that implies 
conservation of total angular momentum in the $z$ direction, $J_z$. It is also 
invariant under Lorentz boost parallel to the $x-y$ plane as we stressed above. 
Thus, in order to label the Hamiltonian eigenstates the conserved quantum 
numbers $E,p_x,p_y$ and $j_z$ (the eigenvalue of $J_z$) can be used. If we boost
the reference frame to obtain $p_x=p_y=0$, the helicity states of the incoming 
plane waves in the unbroken phase correspond to eigenstates of $J_z$ and this 
helicity may be also used to label them.

We define the eigenstates $\psi_n$ by

\beq
H \psi_n=(-1)^r E_n\psi_n,
\eeq

\noindent{where we choose $n=p_x,p_y,j_z,E_n,r$. $r=2,1$ label the positive and
negative energy states, respectively. In [5], for the Hamiltonian}

\beq
H={\vec \alpha}{\vec p}+ \beta m_0 \theta(z),
\eeq

\noindent{
it is stressed that the dimension of the eigenspace is two outside the total 
reflection energy range. In this case the following two set of eigenstates can 
be built}

\bea
\psi_n^{inc}({\vec x})=\left( u_h({\vec p}^{inc}) e^{i{\vec p}^{inc}{\vec x}}+
 R u_h({\vec p}^{inc}) e^{i{\vec p}^{out}{\vec x}} \right) \theta(-z)
\nn   \\
+(1+ R )u_h({\vec p}^{inc}) e^{i{\vec p}^{tr}{\vec x}} \theta(z),\;\;\;
h=j_z.
\eea

\noindent{where $u_h({\vec p}^{inc})$ is a solution of the Dirac equation in the
unbroken phase and the reflection matrix $R$ is given by}

\beq
R={m \gamma^3 \over p_z+p'_z};
\eeq

\noindent{and}

\bea
\psi_n^{br}({\vec x})= \sqrt{{p_z\over p'_z}} \left[ 
\left( u_s({\vec p}^{br}) e^{i{\vec p}^{br}{\vec x}}+
J u_s({\vec p}^{br}) e^{i{\vec p}^{tr}{\vec x}} \right) \theta(z) \right.
\nn   \\
\left.
+(1+ J )u_s({\vec p}^{br}) e^{i{\vec p}^{out}{\vec x}} \theta(-z),
\right]
\eea

\noindent{where $J$ is the reflection matrix when the particle is coming
from the broken phase, given by}

\beq
1+ J ={p'_z\over p_z}(1+ R ),
\eeq

\noindent{and s is a spin index dependent on $j_z$, such that 
$(1+ J )u_s({\vec p}^{br})=u_h({\vec p}^{out})$ is a massless spinor with 
helicity $h=-j_z$; $u_s({\vec p}^{br})$ satisfies the Dirac equation in the 
broken phase. 
The {\cal incoming}, {\cal outgoing}, {\cal transmitted} and {\cal broken 
incoming} four-moment are defined as follows}

\bea
p^{inc}=(E,p_x,p_y,p_z), \nn  \\
p^{out}=(E,p_x,p_y,-p_z), \nn  \\
p^{tr}=(E,p_x,p_y,p'_z), \nn  \\
p^{br}=(E,p_x,p_y,-(p_z)^*). 
\eea

The wave functions $\psi_n^{inc}$ and $\psi_n^{br}$ are orthogonal and allow 
to build an orthonormal basis of the eigenspace beside the analogous 
antiparticle wave functions (for $r=1$ the spinor solution of the Dirac 
equation, $u_h$, must be replaced by the negative energy ones, $v_h$. In what 
follows our conclusions are valid as well for the $r=1$ and $r=2$ eigenstates).
For general walls the argument about the dimension of the eigenspace can be 
generalized and the wave functions obtained to verify

\bea
\psi_n^{inc}({\vec x}) \sim \left\{ \begin{array}{ll}
u_h({\vec p}^{inc}) e^{i{\vec p}^{inc}{\vec x}}+
R u_h({\vec p}^{inc}) e^{i{\vec p}^{out}{\vec x}} 
& -z \gg \delta_w \\
(1+ R )u_h({\vec p}^{inc}) e^{i{\vec p}^{tr}{\vec x}} &
z \gg \delta_w;  \end{array}
\right.
\eea

\noindent{ where $R$ is the reflection matrix in each particular case and
$\delta_w$ defines the characteristic wall thickness as can be seen above.}
Analogously for $\psi_n^{br}$. Thus, $\psi_n^{inc}$ 
describes in the general case an incident plane wave far from the wall
coming from the unbroken phase, bouncing on the wall, and generating a reflected
plane wave in the unbroken phase and a transmitted one in the broken phase.
$\psi_n^{br}$ describes the same process although coming from the broken phase.
In \cite{[7]}, wave functions like $\psi_n^{inc}$ and $\psi_n^{br}$ are obtained
for the particular wall presented in section 5. In this work the authors 
calculate explicitly the overlap integral of these wave functions obtaining a 
non-zero result, concluding that they are not orthogonal. Nevertheless, a 
general argument allows to assert that $\psi_n^{inc}$ and $\psi_n^{br}$ must be 
orthogonal as we will see. 
In \cite{[11]} the error in the overlap integral calculation of
ref. \cite{[7]} is shown and it is explicitly proven 
that the overlap integral is zero for this particular wall profile. 
(the fact that the moment eigenvalues are different in the broken and unbroken phases 
introduce a factor ${p'_z\over p_z}$ which is  forgotten in ref. \cite{[7]}). 
Now we will present the general argument.

The ultimate reason for the orthogonality of the eigenstates considered follows from
the different physical processes to which they are related. The real processes
are described by wave packets coming from the unbroken (broken) phase, 
bouncing on the wall, and generating reflected and transmitted packets. 
The solutions which are asymptotically plane waves have no real physical 
meaning, but they allow to build up localized wave packets. 
As we will see, the solutions named $\psi_n^{inc}$ are associated to wave 
packets evolving from the unbroken phase to the broken phase and those named 
$\psi_n^{br}$ describe the same but in the opposite direction, from broken to
unbroken phase. Since these wave packets do not overlap in the past, by unitarity they will not overlap in all their evolution. Let us give a few more details.

The $x$ and $y$ components of the wave packet are not modified by the wall,
and will be ignored in what follows. We consider the following incoming wave
packet, approaching the wall from the unbroken phase,

\bea
P({\vec p}^{inc},z,t)= 
N \int dk_z e^{-(k_z-p_z^{inc})^2d^2/2}e^{ik_z\tau}u_h({\vec k})
\nn \\
\sim & e^{-(\tau/d)^2}e^{ip_z^{inc}\tau}u_h({\vec p}^{inc}), \label{packet}
\eea

\noindent which has the helicity, $h$, fixed. $N$ is a normalization constant, 
the quantity $\tau=z+Z-t-T$ has been introduced, where $Z/d>>1$ and $T/d>>1$. 
$d$ denotes the spatial extension of the wave packet which is located at the 
time $\sim -T$ around the position $\sim -Z(Z>0)$. $d$ is introduced in such a 
way that $p_z^{inc}d>>1$.
It can be easily proven that at $t \sim -T$,

\beq
P({\vec p}^{inc},z,t)=N'\int dk_zA(k_z,p_z^{inc})\psi_n^{inc}(z),
\eeq

\noindent 
where $N'$ is a new normalization constant and $n=k_x,k_y,j_z,E_k,r$. When 
the terms exponentially suppressed by $e^{-(p_z^{inc}d)^2}$ are 
neglected as in the second line of (\ref{packet}), we obtain

\beq
A(k_z,p_z^{inc})=e^{-(p_z^{inc}-k_z)^2{d^2\over 2}}e^{i(Z-t-T)k_z}.
\label{gaus}
\eeq

\noindent
In other words, an incoming wave packet located far away from the wall in the 
unbroken phase, with group velocity pointing towards the wall, totally expands 
on the eigenstates called {\cal incoming}. In \cite{[5]} this general 
conclusion for any wall profile is stated for the particular case of the thin 
wall (step profile function). Analogous conclusions hold for the wave packets 
coming from the broken phase and for the eigenstates we called {\cal broken 
incoming} ($\psi_n^{br}$).
Obviously the superposition of these two incoming wave packets located far from 
the wall, to the left and to the right, is zero at initial time $-T$. As they evolve 
in the time, for $t=0$, they coincide in the same spatial region but their overlap 
integral keeps on being zero, since the Hamiltonian being hermitian the time 
evolution operator is unitary. It follows that the superposition of the 
two wave packets located in a certain region centered at $z=0$, which extends a 
distance of the order of d, is zero. If the limit $d \rightarrow \infty$
is taken now, the gaussian with
the appropriate normalization constant gives a delta function in (\ref{gaus}), 
and the overlap of the two wave packets becomes the overlap of $\psi_n^{inc}$ and 
$\psi_n^{br}$. Obviously, $d$ and $Z$ go to $\infty$ so that $Z/d \gg 1$,
our argument being valid in this limit. From the vanishing of that overlap 
we conclude to the orthogonality of these eigenfunctions.

\newpage

{\large \bf FOOTNOTES }

\begin{itemize}

\item[$^1$] In these references the final expressions for $R(E)$ and $\overline{R}(E)$ 
are equivalent to (\ref{9}), although the notation is different. Moreover, in these
references $z_0$ is taken to be $0$.

\item[$^2$] This is a reasonable assumption. By considering the 
appropriate dimensions we have that ${\hbar c\over \delta_w}=O[10 \mbox{GeV}]$
(see for example \cite{[7]},\cite{[4]}) and therefore 
$a={2 E \delta_w \over \hbar c}=O[10]$ in the electroweak phase 
transition energy scale. For this range of energy $({E\over m_0})^2 \gg a$ 
except for the top mass.

\item[$^3$] As known the function ${\arctan}(x)$ defined from $\R$
to $(0, 2 \pi)$ is bivalued. The right-hand of the second equation in (\ref{29}) is
well-defined by considering the sign of $(F^{(k)}(1) + F^{(k)}(-1))\sin
a$ to determinate
if the angle belongs to $(0, \pi )$ or to $(\pi , 2 \pi )$. If it is
zero, $(F^{(k)}(1) - F^{(k)}(-1)) \cos a$ must be considered in order to 
determine if the angle is $0$ or $\pi$.

\item[$^4$] In ref. \cite{Rod96b} we use analytic profiles
which behave as $e^{C_-x^{2n+1}}$, $e^{-C_+x^{2n+1}}$, for $x<<-1$,
$x>>1$, respectively. The results which will be derived in what follows
can be, in the same way, obtained for these profiles.

\item[$^5$] The position of the closest pole to the real axis, $y_j$, gives us the 
smallest value of $\lambda$ we can consider in order to be able to neglect the
contribution of $R_{N-C}$.
We have $\lambda \gg {2E\sigma \over C_+} y_j$, ${2E \sigma \over C_-}
y_j$. As it will be seen the term $R_{N-C}$ is associated to the fact of considering the
evolution in the region $(- \delta_w, \delta_w)$ and introducing the technical 
assumption mentioned above.

\item[$^6$] In order to compare results the relationship $a=1/\sigma$, where $a$ is the 
energy parameter used in ref. \cite{[8]}, must be taken into account.   

\end{itemize}


\newpage
\begin{thebibliography}{11}

\bibitem{Rod96} J. Rodr\'{\i}guez--Quintero, M. Lozano and O. P\`{e}ne,
Phys. Lett. {\bf B388}
(1996) 259-265.

\bibitem{[1]} V. A. Kuzmin, V.A. Rubakov and M.E. Shaposhnikov, Phys.
Lett. {\bf B155}
(1985) 36. 

\bibitem{[2]} A.S. Sakharov, JETP Lett. {\bf 6} (1967) 24. 

\bibitem{[3]} M. Kobayashi and T. Maskawa, Prog. Theor. Phys. {\bf 49} (1973) 652. 

\bibitem{[4]}G. R. Farrar and M.E. Shaposhnikov Phys. Rev. {\bf D50} (1994) 774. 

\bibitem{[5]} M. B. Gavela, M. Lozano, J. Orloff, O. P\`ene, Nucl.
Phys. {\bf B430} (1994) 345.

\bibitem{Quimb} M.B. Gavela, P. Hernandez, J. Orloff and O. Pene, Mod.
Phys. Lett. {\bf A9} (1994) 795; M.B. Gavela, P. Hernandez, J.
Orloff, O. Pene and C. Quimbay,
Nucl. Phys. {\bf B430} (1994) 382; Huet and Sather Phys. Rev. {\bf D51} (1995) 379.

\bibitem{[6]} A. E. Nelson, D. B. Kaplan, A.G. Cohen, Nucl. Phys. {\bf
B373} (1992) 453. 

\bibitem{joyce} M. Joyce, T. Prokopec and N. Turok, Phys. Rev. Lett
{\bf 75} (1995) 1695, erratum {\bf 75} (1995) 3375, Phys. Rev. {\bf
D53} (1996) 2930, {\bf D53} (1996) 2958. 

\bibitem{comelli} D. Comelli, M. Pietroni and A. Riotto, Phys. lett.
{\bf B354}  (1995) 91, Phys. Rev. {\bf D53} (1996) 4668. 
 
\bibitem{cline} M. Cline, Kimmo Kainulainen and Axel P. Vischer Phys.
Rev. {\bf D54} (1996) 2451. 

\bibitem{frere} J.M. Frere, L. Houart, J.M. Moreno, J. Orloff
and M. Tytgat, Phys. Lett. {\bf B314} (1993) 289. 

\bibitem{susy} A.G. Cohen and A.E. Nelson Phys. Lett. {\bf B297} (1992) 111. 

\bibitem{[10]} A. E. Nelson, D. B. Kaplan, A. G. Cohen, Phys. Lett.
{\bf B245} (1990) 561;
 \\ A. E. Nelson, D. B. Kaplan, A. G. Cohen, Phys. Lett. {\bf B349} (1991) 727.

\bibitem{pilar} P. Hernandez and N. Rius CERN-TH/96-301, hep-ph: 9611227.

\bibitem{[14]} F. R. Klinkhammer and N.S. Manton, Phys. Rev. {\bf D30} (1984) 2212. 

\bibitem{Hoo76} G.'t Hooft, Phys. Rev. Lett. {\bf 37} (1976) 8; Phys.
Rev. {\bf D14} (1976) 3432; \\ N.S. Manton, Phys. Rev. {\bf D28} (1983) 2019; \\
S. Dimopoulos and L. Susskind, Phys. Rev. {\bf D18} (1978) 4500.  

\bibitem{[7]} A. Ayala, J.J-Marian and L. Mc Lerran, Phys. Rev. {\bf
D49} (1994) 5559. 

\bibitem{[8]} K. Funakubo, A. Kakuto, S. Otsuki, K. Takenaga and F.
Toyoda, Phys. Rev. {\bf D50}
(1994) 1105.

\bibitem{[9]} M. Dine, R. Leigh, P. Huet, A. Linde, Phys. Lett. {\bf
B283} (1992) 219;
Phys. Rev. {\bf D46} (1992) 550 .  

\bibitem{[11]} J. Rodr\'{\i}guez--Quintero, FANSE-96/18 (1996) report
to obtain the {\bf PhD} degree. 

\bibitem{[12]}  I.S. Gradshtein and I.M. Ryzhik, {\cal Tables of Integrals, 
Series, and Products} (Academic Press Inc. 1965). 

\bibitem{Tor96} E. Torrente-Lujan, hep-ph/9607335 (1996).

\bibitem{Rod96b} J. Rodr\'{\i}guez--Quintero,
FANSE-96/10--hep-ph/9610290 (1996), to be published in Phys. Lett. {\bf B}.

\bibitem{Schi68} See for instance L. Schift, {\cal Quantum mechanichs} (McGraw-Hill 
Inc. 1968); G. Gamow, Z.Physik {\bf 51}, 204 (1928); R.W. Gurney and E.U. Condon, 
Phys. Rev. {\bf 33}, 127 (1929).


\end{thebibliography}
\end{document}